# Intermixed Cation-Anion Aqueous Battery Based on an Extremely Fast and Long-Cycling Di-Block Bipyridinium-Naphthalene Diimide Oligomer

Sofia Perticarari[a,b], Tom Doizy[a], Patrick Soudan[a], Chris Ewels[a], Camille Latouche[a], Dominique Guyomard[a], Fabrice Odobel[b]*, Philippe Poizot[a,c]* and Joel Gaubicher[a]*

a-  Institut des Matériaux Jean Rouxel (IMN), Université de Nantes, CNRS, 2 rue de la Houssinière, B.P. 32229, 44322 Nantes Cedex 3, France

E-mail: Joel.Gaubicher@cnrs-imn.fr and Philippe.Poizot@cnrs-imn.fr,

b-  CEISAM, Chimie et Interdisciplinarité, Synthèse, Analyse, Modélisation, Université de Nantes, 2 rue de la Houssinière, B.P. 92208, 44322 Nantes Cedex 3, France

E-mail: Fabrice.Odobel@univ-nantes.fr

c-  Prof. Philippe Poizot
    Institut Universitaire de France (IUF), 103 bd Saint-Michel, 75005 Paris Cedex 5, France

**Abstract:**

Aqueous batteries, particularly those integrating organic active materials functioning in a neutral pH environment, stand out as highly promising contenders in the stationary electrochemical storage domain, owing to their unparalleled safety, sustainability and low-cost materials. Herein, a novel di-block oligomer (**DNVBr**), serving as the negative electrode of an all-organic aqueous battery, is shown to offer exceptional output capabilities. The battery's performance is further enhanced by a unique intermixed p/n-type storage mechanism, which is able to simultaneously exchange light and naturally abundant $Na^+$, $Mg^{2+}$ and $Cl^-$. Reaching up to 105 mAh/g, this system shows remarkable capacity retention for several thousand cycles (6500 cycles, ~40 days) in various neutral electrolytes, including raw ocean water (~3000 cycles, ~75 days). The surprisingly fast kinetics of this di-block oligomer allow to attain an unmatched specific capacity of near to 60mAh/$g_{electrode}$ while entirely devoid of conducting additives, and more than 80mAh/$g_{electrode}$ with 10% carbon additive, as well as displaying an areal capacity as high as 3.4mAh/$cm^2$ at C rate. Full cell validation was demonstrated over 1600 cycles by virtue of a commercial TEMPO molecule, which permitted an energy density of close to 40Wh/$kg_{materials}$ at C rate in a self-pH-buffered and inexpensive aqueous electrolyte.

**Introduction**

By virtue of their high energy efficiency, facile scalability, intrinsic safety, not to mention low cost and environmental compatibility, aqueous batteries are being increasingly considered as a novel and promising technology for grid storage[1–3]. Despite recent progress,[1,4–12] however, these systems are still associated with a relatively low energy density of below 50Wh/$kg_{cell}$, or insufficient cycle life[12], which is detrimental to their economic viability by perpetuating the issues of both unit cost and/or durability of active materials. A device of this sort needs to be able to supply an adequate amount of cycles over a sufficient time span in order to equalize energy storage costs to around 100 $/kWh. Furthermore, the extensive utilization of batteries for both domestic and large-scale applications that would result from the global energy transition, would further restrict our already limited, unequally distributed and monopolized metal resources. The design of suitable host materials displaying optimal potential and high chemical/electrochemical stability hinges upon resolving these challenging issues.

In this regard, substituting transition metal-based materials with organic ones appears to be a very favorable approach[1,2] since these materials provide several distinct advantages: *(i)* their extraction is not restricted to specific geographical areas, making them of particular interest in countries such as Europe where certain key raw materials are scarce; *(ii)* low-cost organic chemistry can provide a virtually infinite number of compound modifications with appropriate functional groups (hydrophobic, electron attractor/donor, favorable pi-stacking interactions, etc.), which permits the tuning of their solubility, molecular mass and chemical reversibility, as well as the adjustment of their redox potential, thereby optimizing the cell voltage without triggering overwhelming $O_2$ or $H_2$ evolutions. By way of reminder, in an electroactive organic group, the latter property arises from the change in its charge state, whereas for inorganics it stems from the change in the valence of the transition-metal or element; *(iii)* they can undergo multi-electron redox reactions which can lead to a much higher specific capacity. Lastly, it is worth noting that their low volumetric density is evidently not as great a hindrance with respect to stationary applications as it is to itinerant ones.

From a general perspective, and based on several exhaustive studies reviewing organic redox materials[1,2,13–19] it is clear that although many n-type organic materials such as quinones and diimides can be designed for the anode side fairly "easily" (potentials below -0.3 V vs. Saturated Calomel Electrode (SCE) or 2.95 V vs. Li$^+$/Li) due to the carbonyl/enolate redox moiety, it is much more challenging to find cathodic n-type materials with sufficient potentials to produce a voltage of >1 V. On the other hand, while the redox potentials of the viologen group match up with the negative side, the potentials of other p-type organic materials are a fit for the positive side. It is for this reason that cationic (n-type) rocking-chair aqueous batteries only make use of hybrid cells that combine inorganics (cathode) and organics (anode). Yao's group[4] aptly illustrate this point with the use of a polypyrene-4,5,9,10-tetraone negative electrode (220 mAh/g$_{material}$) paired with LiMn$_2$O$_4$. This full cell can sustain ~90 Wh/kg$_{materials}$ for more than 3000 cycles at 0.23 A/g (3500 h cycling) with near to 100% coulombic efficiency. Another interesting hybrid cationic cell was demonstrated by Wu and co-workers [20] using an ammonium Ni-based Prussian white as the cathode paired with a 3,4,9,10-perylenetetracarboxylic diimide in 1 M (NH$_4$)$_2$SO$_4$. The cell delivered 43Wh/kg$_{materials}$ at 1.5C, but displayed a rather moderate capacity retention (67% upon 1000 cycles at 3C rate). Thus far, the only full organic *rocking-chair* aqueous cells are p-type ones (anionic configuration), and both of them were put forward by the same group (Nishide, Oyaizu and co-workers): their thin film batteries were designed using a TEMPO redox polymer derivative, with poly (2,2,6,6-tetramethylpiperidin-4-yl) acrylamide (PTMA) as the cathode, paired with either a highly cross-linked polyviologen hydrogel (poly-(tripyridiniomesitylene))[21] enabling ~1.3 V over 2000 cycles, or with poly(*N*-4,4'-bipyridinium-*N*-decamethylene dibromide) leading to 2000 cycles with a 1.2 V average voltage [22]. Recently, Truhlar, Wang and co-workers demonstrated a full organic "*dual-ion*" cell (cations and anions are drawn from the electrolyte, as opposed to a "rocking-chair" system) based on p-type polytriphenylamine and n-type polynaphthalene diimide polymers as the positive and negative materials, respectively[9]. The authors determined that nearly 53 Wh/kg$_{materials}$ and 32 kW/kg$_{materials}$ can be obtained for 1 mg/cm$^2$ electrodes using a 21 m LiTFSI water-in-salt electrolyte. Lastly, we previously reported a possible new avenue for designing aqueous batteries using an organic material wherein a p-type viologen and n-type naphthalene diimide moieties merge together into a short oligomer, thus allowing the exchange of both anions and cations in a narrow potential range, and showing encouraging performance[23]. In light of the above, even though remarkable progress has already been made, further material innovation is required in order to ultimately obtain a low-cost, green and long-lasting aqueous battery for renewable energy storage.

In the present work we identify an advanced p/n-type organic scaffold that exchanges simultaneously cations and anions in a high cation to anion ratio and over almost all of its potential range, and with greatly improved capacity (up to 82 mAh/g$_{electrode}$), cyclability (over 6500 cycles), chemical stability (3000 cycles in ocean water) and conductivity, thereby bringing us quite a bit closer to being able to build competitive devices. This point was demonstrated by pairing a commercial TEMPO small molecule for use as the cathode material, thus attaining up to 40 Wh/kg$_{materials}$ at a C rate.

RESULTS

**Synthesis**

In order to attain the objective of water-insoluble p/n-type redox active compounds providing optimal performance, a new **"di-block"** oligomer (**DNVBr**, Scheme 1) was designed. The latter contains naphthalene diimide repeating units (referred to as NDI) coupled to bipyridinium units (referred to as Violo) by a propyl linker. As mentioned in the introduction, being able to synthesize active materials at a low cost constitutes a *sine qua non* condition for making ion-aqueous batteries, a technology of choice for stationary applications. In this spirit, **DNVBr** was synthesized by following two elementary reaction steps (Scheme1): firstly, the intermediate NDI **3** was obtained, in a yield of 89%, by an imidization reaction between the naphthalenetetracarboxylic dianhydride NDA (**1**) and the 3-bromopropylammonium bromide (**2**), according to the customary procedure[24]. The second step was a nucleophilic substitution between compound **3** and the 4,4'-bypiridine. The final dark brown product (**DNVBr** in Scheme 1) is completely insoluble in most organic solvents, thus preventing its characterization by Size-Exclusion Chromatography and by Mass Spectrometry analysis, thereby precluding the precise determination of this new material's polydispersity. However, thanks to its solubility in trifluoroacetic acid (TFA), the $^1$H-NMR and $^{13}$C-NMR spectra were successfully recorded (Figures S1-S2), showing that a polydisperse mixture of oligomers was produced. The degree of oligomerization was deduced by the integration of the specific signals into the $^1$H-NMR spectrum, and led to an estimated average degree of oligomerization equal to 3±1 (for further details see Figure S1 and accompanying explanations). This result was replicated across several reaction batches and by different experimentalists.

FT-IR/ATR analyses confirm the presence of both the imide and the bipyridinium units *via* the existence of the characteristic stretching bands of the C=O (1702, 1656 cm$^{-1}$), C=N$^+$ (1640 cm$^{-1}$) and C=N (1577 cm$^{-1}$) bonds (Figure S3). The thermal stability of the new product was tested by means of TGA-MS/DSC analysis, showing that **DNVBr** is stable up to 300°C and contains approximately 5wt% adsorbed water (Figure S4). In addition, the FT-IR spectrum indicates a large band at 3400 cm$^{-1}$ that may correspond to the solvation sphere around the Br$^-$ counter-anion.

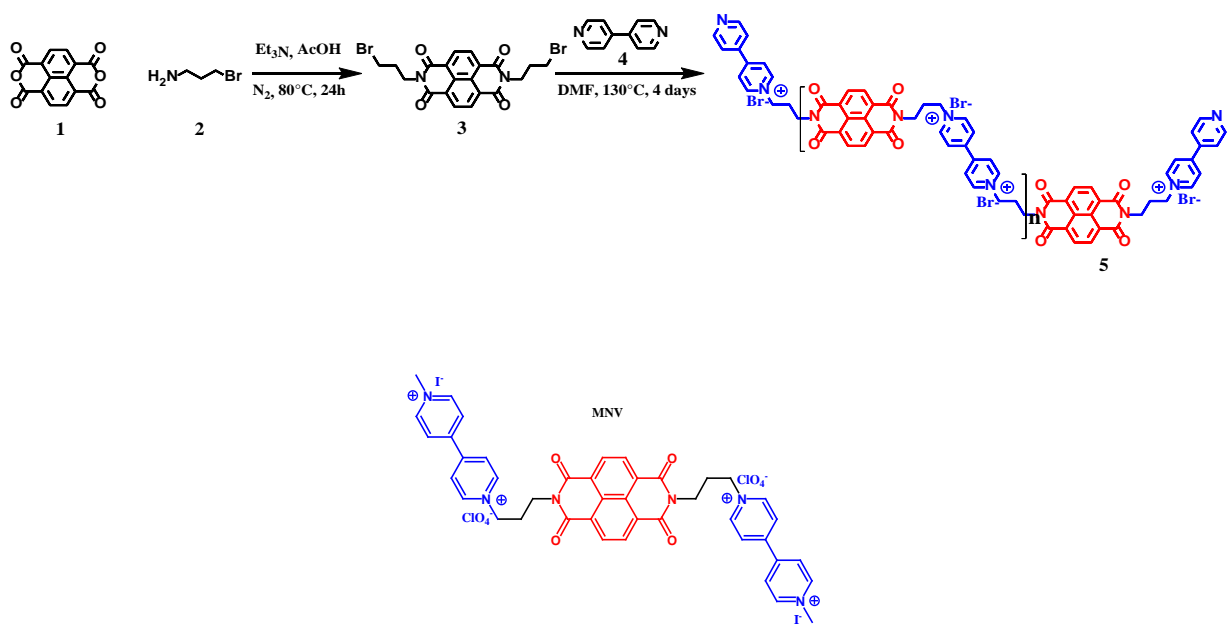

**Scheme 1:** Synthetic route to compound DNVBr having n = 3±1. The structure of MNV is shown for the sake of comparison.

**Electrochemical behavior of DNVBr:**

The **"di-block"** nature of **DNVBr** is characterized by two redox units: the viologen one, ensuring anion exchange (p type); and the NDI one, retaining/releasing cations (n type) according to the following reactions:

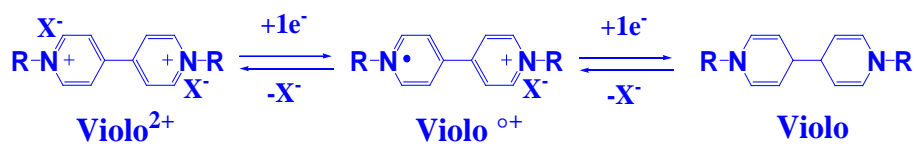

and

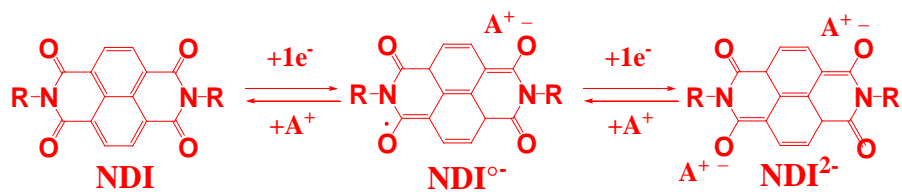

**Scheme 2: Reversible one-electron reduction steps of both viologen (Violo) and naphthalene diimides (NDI) redox-active moieties in blue and red, respectively.**

Importantly for the following results, the capacity of **DNVBr** can in principle be increased with the further reduction of both (Violo$^{\bullet+}$) and (NDI$^{\bullet-}$) to the neutral Violo and dianion (NDI$^{2-}$):

Noteworthy, the second electron reduction of Violo$^{\bullet+}$ could not be achieved, even with a potential as low as -1.V vs. SCE in an aqueous electrolyte. The theoretical specific capacity of the Viologen moiety substituted by a propyl group on one side and bearing Br$^-$ as the counter anions, ($M_w$=374.12 g/mol), is 74.6 mAh/g for a one-electron reaction. However, the capacity of the naphthalenediimide, also functionalized by a propyl linker on one side, is 16% higher (86.7 mAh/g) for one electron and 132% higher for two electrons (173.3 mAh/g). Accordingly, taking into account the side NDI unit and two non-electroactive pyridine end groups, the theoretical specific capacity of **DNVBr** increases asymptotically with n because the NDI/Violo molar ratio decreases (Figure S5). This serves to demonstrate the great potentiality of this kind of di-block assembly that can in principle achieves nearly 120 mAh/g and 160 mAh/g depending on the electrochemistry of the Viologen unit. Second, for any molecular weight (n≥1) the specific capacity of **DNVBr** surpasses that of MNV. Lastly, for n=3 $Q_{DNVBr}$=104.8 mAh/g (11e$^-$), which matches the experimental value achieves at low C rate (C/5) by **DNVBr** (107 mAh/g, Figure S6). This result therefore confirms the narrow polydispersity (~n=3) of **DNVBr** as inferred from the chemical analysis of the compound.

As will be demonstrated throughout this paper, the extended length and specific molecular structure of **DNVBr** offer us numerous advantages over the first di-block compound of this kind[23] (referred to as MNV), making it one of the most attractive negative electrode materials for aqueous batteries to date: **DNVBr** was found to (i) support light counter anions (Br$^-$ and Cl$^-$) without dissolving in any of the aqueous electrolytes (ii) display unmatched performance without carbon additive, (iii) show an intermixed cation-anion mechanism over almost all of its potential range, and (iv) demonstrate outstanding cyclability.

The cyclic voltammogram (CVs) of MNV and **DNVBr** composite electrodes in NaClO$_4$ 2.5 M are shown in Figure 1a,b. The electrochemical profile of both compounds is characterized by three main reversible peaks (denoted by I, II and III, respectively) that are much broader in the case of **DNVBr**. We note that the peak potentials of II could only be distinctively identified upon oxidation for **DNVBr** and upon reduction for MNV, by limiting the potential cut-off (Figure 1a,b). As shown in[23] by testing Di-Methyl-NDI and Di-Methyl Viologen in NaClO$_4$ 2.5 M, peak I and II of **DNVBr** are most likely associated with the NDI and Viologen redox centers respectively. Under these conditions, the estimated equilibrium potential for II$_{DNVBr}$ (approximated to $E_i = 1/2(E_i^{peak,ox} + E_i^{peak,Red})$) is 20 mV lower than for MNV (Table S1). Using the same approximation, the potential differences between I$_{DNVBr}$ and I$_{MNV}$ on one side, and III$_{DNVBr}$ and III$_{MNV}$ on the other side, are +30 and +60 mV respectively (Table S1). Lastly, **DNVBr** is characterized by lower polarizations for I and III. Most especially, the maximum intensity of III$^{red}_{MNV}$ is at −0.901 V, which is 107 mV lower than that of DNV (−0.794 V). Described in more detail below, these changes in equilibrium potential and polarization give **DNVBr** a decisive advantage over MNV as regards its specific capacity.

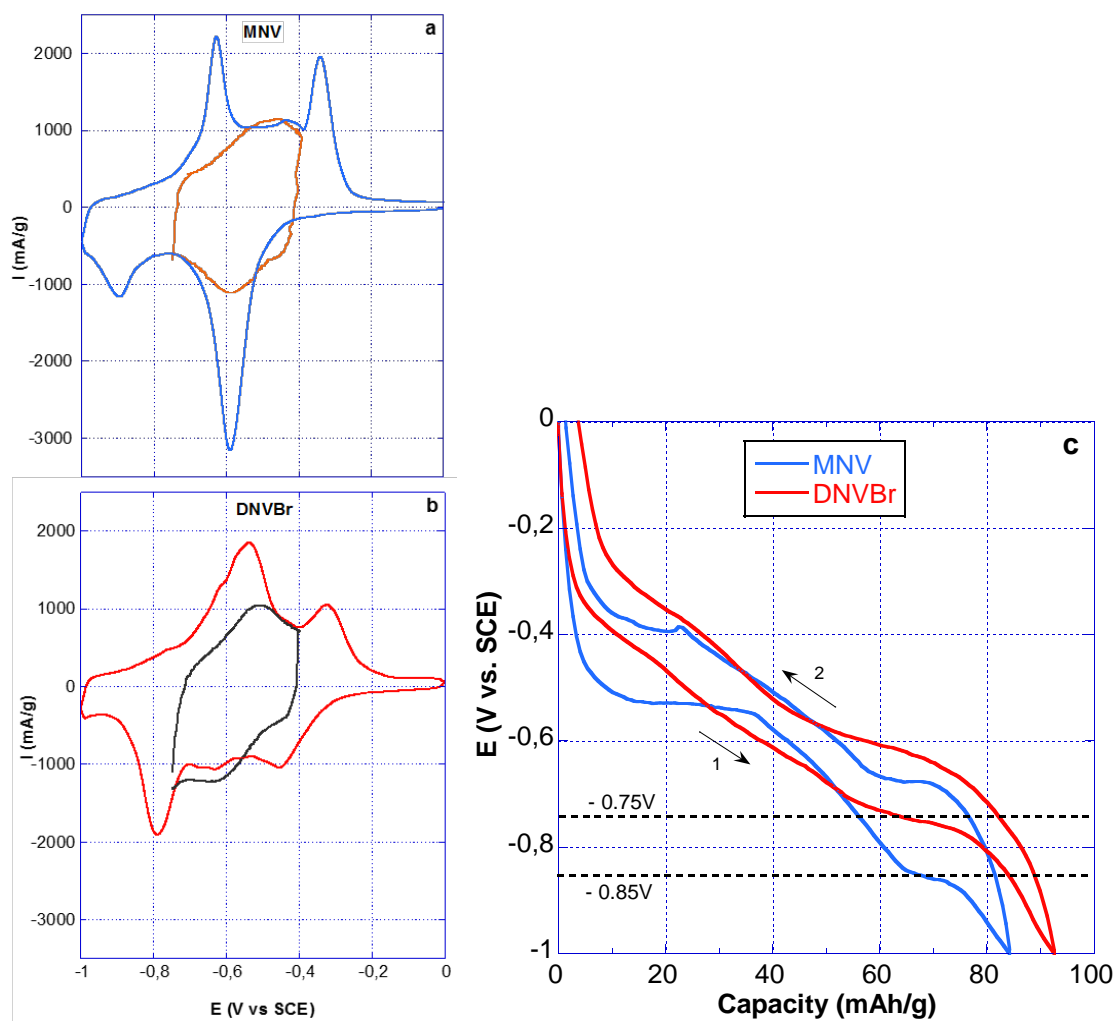

**Figure 1:** (a) Typical cyclic voltammogram of DNVBr (blue) and MNV (red) composite electrodes measured in NaClO$_4$ 2.5 M at a scan rate of 2 mV/s between 0 and -1 V vs. SCE. Limitation of the potentials (pink and light blue) was performed to enable a better description of the electrochemical reaction, denoted as II. (b) Comparison of DNVBr and MNV galvanostatic charge discharge profiles at 0.3 A/g (4C rate) in NaClO$_4$ 2.5 M between 0 and -1 V vs. SCE.

Galvanostatic tests of **DNVBr** and MNV at 0.3 A/g (4C rate, Figure 1c) highlight the fact that moving from MNV to **DNVBr** has a profound effect on the electrochemical profile. In particular, the two potential plateaus that are observed for MNV mirror the occurrence of phase transformations[23], whereas the smooth profile of **DNVBr** suggests solid solution type processes. Reduction of these materials down to −1 V at 4C rate permits to recover a specific capacity on oxidation of 89 mAh/g for **DNVBr**, and 83 mAh/g for MNV (Figure 1c). However, in order to mitigate hydrogen evolution in neutral and molar range electrolytes, such a low cut-off potential can only be used at high current loads[23] (typically 2.4 A/g - 32C-rate). At lower currents, long cycling is typically performed down to -0.75 or even -0.85 V. Figure 1c shows that with a -0.85V cut-off **DNVBr** achieves 85 mAh/g at 4C rate which is +36% higher than MNV. This gain results from both the 60 mV higher average potential and lower polarization of step III.

**Specific insertion mechanism of DNVBr:**

The specificities of the **DNVBr** insertion mechanism were pinpointed by UV-Vis spectroelectrochemistry, EQCM, *operando* synchrotron XRD, as well as spin-polarized DFT calculations.

An overview of the UV-Vis spectral changes observed while cycling **DNVBr** from 0 to −1.1 V vs. SCE by CV (Figure 2a) is shown in Figure 2b, while the evolution of the different redox forms is reported in Figure 2c. The latter is shown by plotting the relative intensity of the specific bands for each species as a function of the scan number. For the sake of clarity, the evolution of the five most representative wavelengths are reported in Figure 2c.

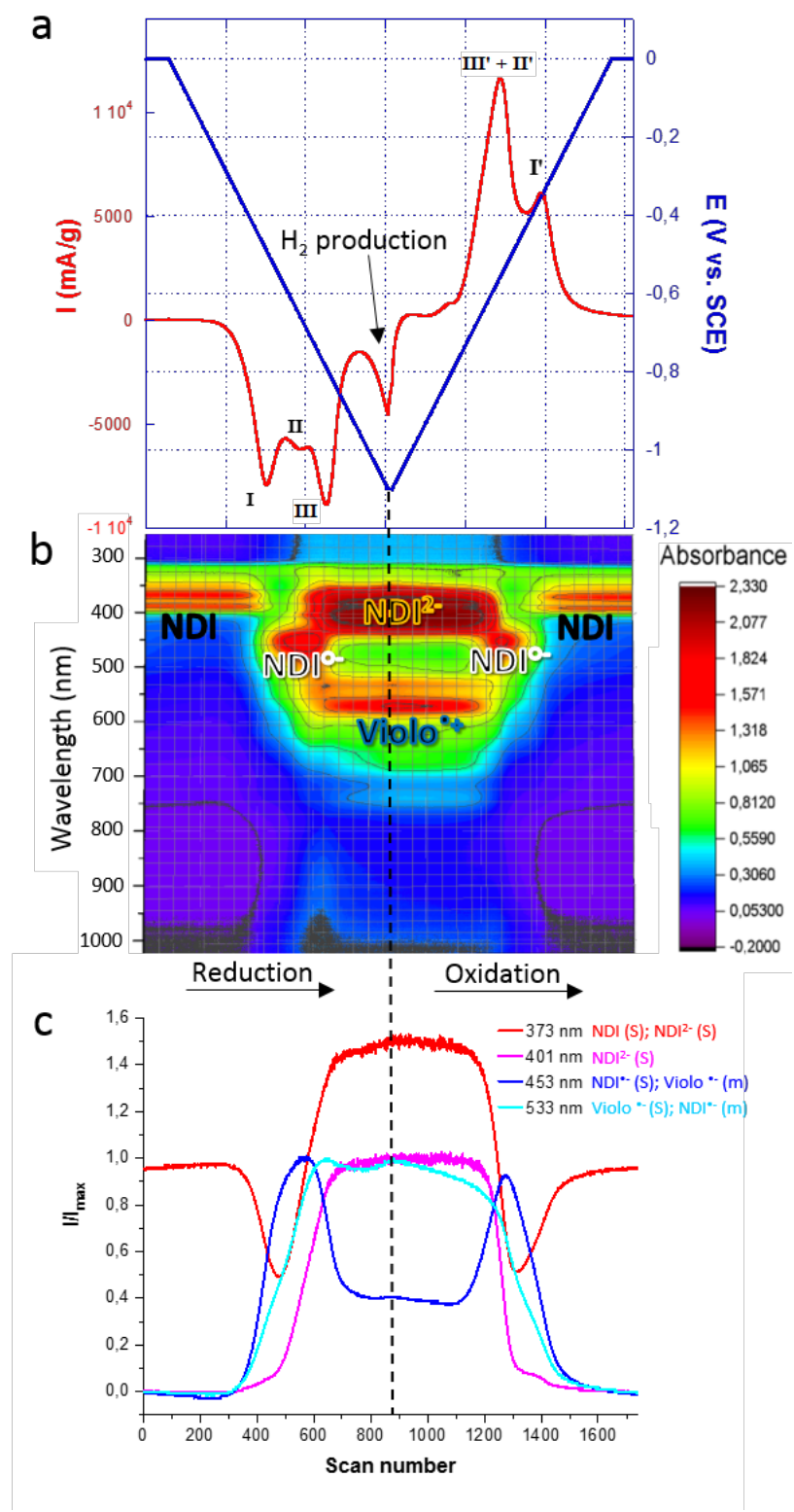

**Figure 2:** (a) CVs of DNVBr measured in NaClO$_4$ 2.5 M at a scan rate of 0.5 mV/s, (b) corresponding UV-vis spectroelectrochemical spectra (most intense components being labelled), and (c) corresponding evolution of the NDI, NDI$^{•-}$, NDI$^{2-}$ and Violo$^{•+}$ related proportions ("S" and "m" referred to as Strong and medium intensity respectively). $I_{max}$ is associated with the maximum intensity observed for a given band during the entire reduction-oxidation cycle. For the sake of clarity, as regards the red curve, $I_{max}$ was arbitrarily set to the intensity of the NDI band in the initial state ($I_0$) of DNVBr.

| Wavelength (nm) | Species and relative intensity | Reference |
|---|---|---|
| **373** | NDI (strong) + NDI$^{2-}$ (strong) | 25 |
| **401** | NDI$^{2-}$ (strong) | 25 |
| **453** | NDI$^{•-}$ (strong) + Violo$^{•+}$ (medium) | 25,26 |
| **533** | Violo$^{•+}$ (strong) + NDI$^{•-}$ (medium) | 25,26 |

**Table 1:** Attribution of the most representative UV-Vis absorption bands relative to the reduction and oxidation of DNVBr in NaClO$_4$ 2.5M.

The first drop in intensity of the red curve associated with the neutral NDI (372 nm) occurs during scan 280 ($E = -0.21$ V), at the expense of the NDI radical anion at 452 nm (dark blue curve). This transformation, which is characterized by an isosbestic point (Figure S7), ends at scan 480 ($E = -0.51$ V) and thereby accounts for the first reduction peak (I) in the CV plot (Figure 2a). It is worth noting, however, that the maximum intensity of the dark blue curve associated with the NDI radical anion keeps increasing until scan 532 (Figure 2c), which corresponds to $E = -0.60$ V in Figure 2a. This demonstrates that a second component is at play at this wavelength. Based on the literature findings[26], this corresponds to the radical viologen moiety (Table 1). The behavior of the viologen redox center can also be tracked by the band at 533 nm (light blue curve, Figure 2c), which grows on top of that pertaining to the radical NDI$^{•-}$ (Figure 2c, Table 1 and Figure S7). From scan 314 on ($E = -0.26$V in Figure 2a) the intensity of this band increases due to the formation of NDI$^{•-}$, but an inflexion point occurs at scan 390 ($E = -0.39$V in Figure 2a) which mirrors the appearance of the viologen cation radical. This intermixing of both the NDI$^{•-}$ and Violo$^{•+}$ is further supported in Figure 2c by the opposite slope changes of the light and dark blue curves in the vicinity of scan 480 ($E = -0.51$ V in Figure 2a): *(i)* the negative inflexion of the dark blue curve corresponds to the end of the NDI to NDI$^{•-}$ transformation whereas, *(ii)* the positive inflexion of the light blue curve is associated with the maximum of intensity of the CV peak II in Figure 2c. Consistently, this CV peak can confidently be ascribed to the reduction of Violo$^{2+}$ to Violo$^{•+}$. Finally, the band at 452 nm (dark blue curve) decreases during the third electrochemical process due to the reduction of NDI$^{•-}$ into its dianionic quinoid form NDI$^{2-}$. The appearance of the NDI dianion can indeed be identified from the band at 401 nm (Figure 2c, pink curve), for which the slope changes from scan 493 ($E = -0.53$ V in Figure 2a). Notably, this characteristic is therefore unique to **DNVBr**, since the dianionic quinoid form $NDI^{2-}$ was not detected for MNV even at potentials as low as -0.75V[23]. Taking the $NDI/NDI^{\bullet-}$ redox couple as an example, this transformation is characterized by an isosbestic point (Figure S7) confirming that these redox centers are not involved in side reactions. The production of $NDI^{2-}$ can also be detected at 372 nm (Table 1) as the intensity of the red curve increases above $I/I_{max}(I/I_0)=1$ from the same scan number (Figure 2c). The two $NDI^{2-}$ bands stop growing as the intensity of the low-potential CV peak (III, in Figure 2a) tends toward zero in the vicinity of scan 800 ($E = -0.98$ V). The subsequent increase in reduction current at $\sim-0.92$ V (scan 780, Figure 2a) is therefore mainly attributed to $H_2$ evolution. Lastly, one can see that the intensity of the two blue curves, as well as the red and pink ones, increases slightly from scan 780 up until the end of the reduction. Accordingly, it appears that *(i)* both the $NDI^{\bullet-}$ and the viologen subunits are still active until the very end of the reduction, and *(ii)* further reduction of the viologen radical cation into its neutral form does not occur under our conditions, which is an important feature considering that the neutral form is known to induce considerable volume variations and poor reversibility[26]. Upon reverse scan, the oxidation of the electro-generated species reverts to the initial spectra. An asymmetry appears between the end of reduction (from scan 650 to 900) and the beginning of oxidation (scan 900 to 1250). This asymmetry, comparable to the one observed in the electrochemical curve, is characterized by two features: *(i)* the drop of intensity in the radical viologen bands (dark and light blue curves) is more pronounced at the beginning of the oxidation process than the corresponding increase in intensity at the end of the reduction, and *(ii)* the $NDI^{2-}$ starts decreasing upon oxidation around 50-100 scans later than one would expect judging by its behavior upon reduction.

Overall, these results demonstrate that **DNVBr** shows an intermixed p-n type electroactivity over nearly its entire potential range, a property which has never yet been encountered in the battery field to our knowledge. This particularity, which stems primarily from the stabilization of its dianionic quinoid units ($NDI^{2-}$) to high potentials, accounts for the high specific capacity of **DNVBr**. It is anticipated that such an extended anionic-cationic ingress-release in opposite flows (see EQCM results below) could potentially mitigate volume variations in the electrode, which would be an even greater advantage since thick ones are mandatory for decreasing levelized energy costs.

It is worth noting that, a very similar evolution of the UV-Vis response is observed when using $Mg(ClO_4)_2$ 1.25 M as the electrolyte (Figure S8).

The electrogravimetric behavior of **DNVBr** measured by EQCM (Figure S9) provides valuable information supporting the dual cationic/anionic insertion processes of **DNVBr**. Indeed, on reduction three successive steps are observed: a first gain in mass that can be attributed to the uptake of cationic species occurs upon formation of the radical anion ($NDI^{\bullet-}$) during the electrochemical step (I); this process is followed by a mass loss, consistent with the reduction of the p-type Viologen to its cationic form ($Violo^{\bullet+}$) during (II) and the concomitant release of anions; finally, the low potential step (III) is associated with a second gain in mass that supports the reduction of the radical anion ($NDI^{\bullet-}$) to its dianionic form $NDI^{2-}$. Therefore, the EQCM finding is in agreement with the UV-Vis inferred description of electrochemical behavior. Given the intricate intermixing of the p/n-type electrochemical processes, a qualitative analysis would, however, require a dedicated study and is thus beyond the scope of this paper. Nevertheless, based on the present results, calculations suggest an important co-insertion of water molecules during cation ingress: eighteen $H_2O$ molecules per cation during step I, and thirteen during III; while during II, the departure of one $ClO_4^-$ anion would be counterbalanced by the ingress of two $H_2O$ molecules (details are reported in Figure S9).

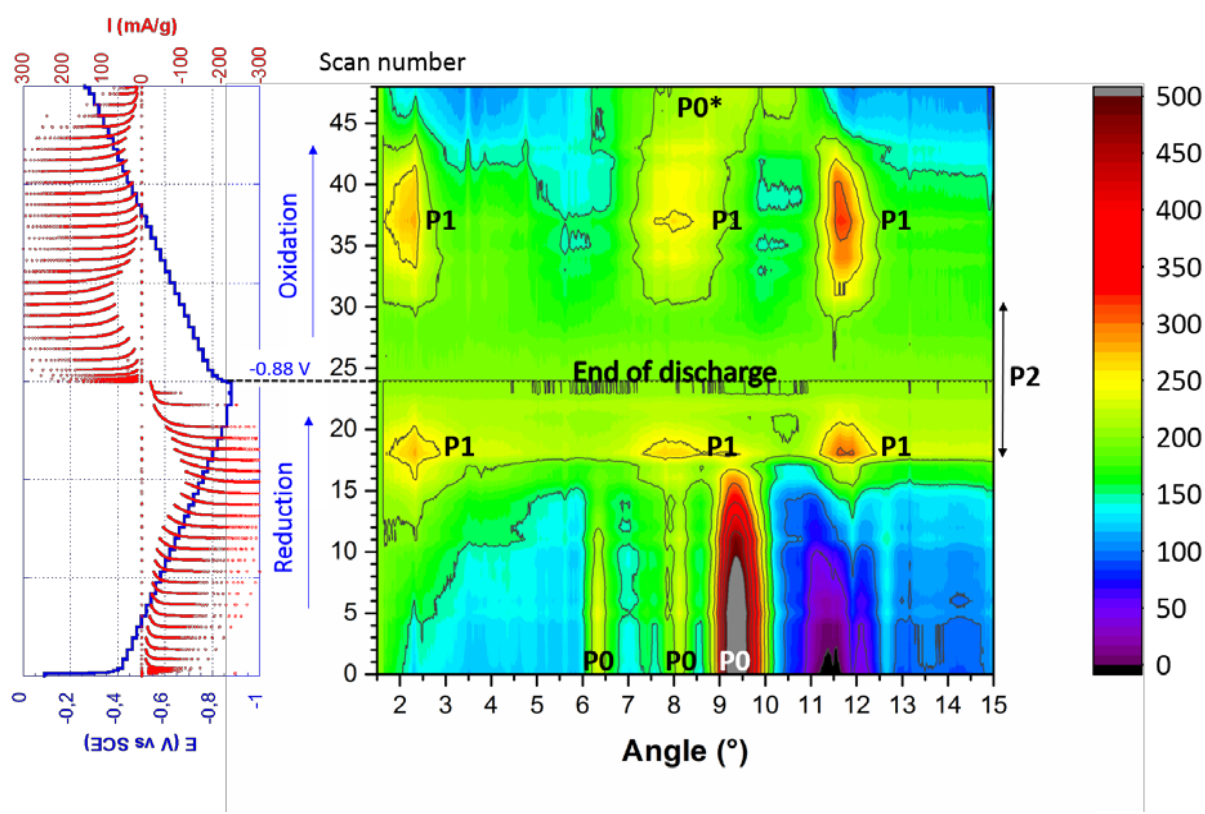

**Figure 3:** Evolution of XRD diagrams of **DNVBr** upon subtraction of the background during 1st reduction (scan 0-24) and oxidation (scans 25-48). The double arrow on the right-hand side represents the domain associated with the existence of P2.

The structural evolution of solid state oligomers or polymers during electrochemical ion insertion is certainly a challenging issue given the poor crystallinity of this class of materials. Synchrotron XRD, however, reveals fairly unexpected behavior for **DNVBr** (Figure 3). Indeed, although the smooth electrochemical profile of **DNVBr** suggests solid-solution type processes, no peak shifts were detected during the entire electrochemical process (reduction and oxidation, Figure 3). Rather, the intensity of **DNVBr** peaks (referred to as phase P0) at 6.35°, 8.15°, 9.40° and 12.55° starts decreasing as soon as the reduction begins (scan 5). This is most clearly visible for the main peaks at 6.35°, 8.15° and 9.40° in Figure S10a. Disappearance of P0 occurs upon the appearance of a a new set of peaks at 2.33°, 8.03° and 11.79° (the corresponding phase being referred to as P1). The P0 to P1 phase transformation is completed upon scan 18 ($Q-Q_0$

= 61.5 mAh/g). It is worth noting that the new peak at 2.33° corresponds to a very large $d$-spacing ($d$ =33 Å), which could either be due to a superstructure or the uptake of a large amount of water molecules, as suggested by EQCM analysis. From scan 18 up until the end of the reduction (scan 24, $Q$–$\underline{Q}_0$ = 81.4 mAh/g), P1 peaks decrease ((Figure 3). However, a thorough examination of the data did not reveal the emergence of a new set of peaks, leaving us to conclude that an amorphous compound is formed (referred to as P2). This amorphization-type transformation is tentatively ascribed to further water uptake and a swelling of the structure during Na insertion and concomitant formation of NDI$^{2-}$. Unlike during the reduction step, where the two phase transformations are merged into a single large electrochemical envelop (Figure 3), the oxidation is characterized by two main peaks I and (II+III) as previously described in Figures 1 and 2. Remarkably, P1 re-forms during the (II+III) oxidation process and reaches a maximum when the current response drops to nearly zero at scan 37. As expected from Figure 1 and 2, the corresponding capacity is ($Q$–$Q_{red}$) = 48.4 mAh/g, which is 66% of the total capacity upon oxidation (73.0 mAh/g). During step I, a second phase transformation characterized by the full disappearance of P1 occurs upon the appearance of a very broad peak from 6° to 11° (the latter can be better observed in Figure S10b). The angular range of this peak encompasses the initial **DNVBr** peaks, and therefore suggests that the corresponding re-oxidized phase (referred to as P0*) resembles **DNVBr**. This intriguing contradiction between the smooth potential behavior (Figure 1) that suggests a solid solution type structural conduct, and the formation of new structural phases, has never been reported to our knowledge. As shown by the XRD diagrams in Figure S11, **DNVBr** is much less crystalline than MNV, which indicates a phase transformation associated with a potential plateau[12]. The **DNVBr** molecule, which is nearly three times longer, is expected to inhibit long-range ordering and favor crystalline strains, this effect being presumably more pronounced at the surface of the grains. In this possible scenario, the potential being a surface measurement, we tentatively propose that the relatively less disordered inner part of the grains would favor the formation of the new phases, thereby leaving the surface with a wide variety of ionic and electronic states, and therefore a sloppy profile throughout the redox process.

To further support the unique properties of **DNVBr**, we used local spin density functional theory to model DNV and MNV as isolated solvent-free molecules. Both systems were fully geometrically optimized, and with the molecules in the neutral charge state (i.e. with full charges received from counter-anions, which are not explicitly included). In agreement with the experiment, both molecules are initially polarized with charge transfer from the Violo to the NDI. In MNV, 0.5e is donated from Violo to NDI (0.25e from each of the two Violo groups), whereas in DNV the ratio of NDI to Violo groups is different and as a result the Violo groups are more positive (+0.43e on average) and the NDI less negative (-0.44e).

Figure 4 shows the calculated Kohn-Sham electron eigenvalues for the DNV molecule. The molecule has a wide 3.3eV separation of filled and empty levels with nine additional states (labeled a-i in Figure 4) with nearly degenerate energy at the Fermi level. These mid-gap levels contain eight electrons in the neutral species.

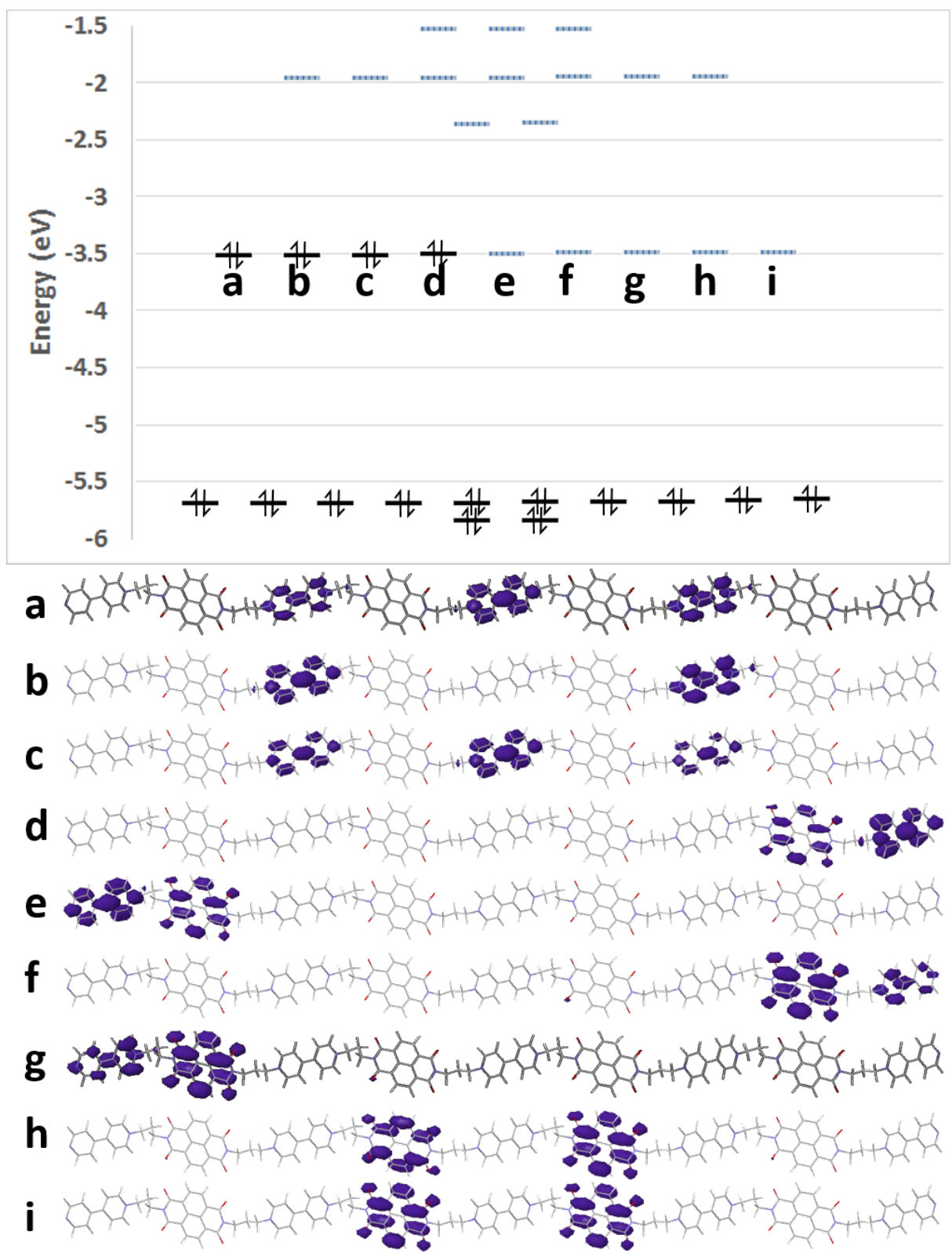

Figure 4: Calculated LDA-DFT Kohn-Sham eigenvalues (eV) for an isolated DNV molecule, showing the real-space wave function distribution on the molecule for states at the Fermi level. Level offset is for visual clarity and does not represent symmetry equivalent states. The molecule is in the neutral charge state, *i.e.* after charge transfer from nominal counter-anions, which are not included explicitly in the calculation.

In the experimental crystal environment, the degeneracy of these states will be lifted by the presence of the counter anions. Since these will be distributed stochastically amongst available interstitial sites in the crystal, we expect the ordering of these levels a-i to vary between DNV molecules, which explains the redox peak broadening seen in experiment compared to the previous MNV studies. Since the system initially contains only anions, in general these will tend to localize near the positive Violo groups, stabilizing states a-c and pushing up empty states d-i. The first states to populate during electron addition will thus be these lowest unoccupied states d-i on the NDI. Once these molecules are charged, further electron addition to the sample will then be to the remaining DNV molecules with empty Violo states near the Fermi level (a-c).

Importantly, as charge is added to the DNV molecule these states split further (see Figure S12), while we expect the counter-ions to redistribute in response to charge filling. Furthermore, as electrons are added these gap states are driven upwards in energy and eventually encounter the higher energy empty states above them, which are more delocalized between Violo and NDI groups (Figure S12). This is consistent with the observed increase in sample conductivity at lower potentials (see below). Most interestingly this effect could corroborate the intermixed Violo-NDI character upon electron filling.

Additionally, the gap states in DNV are distributed across multiple groups of the same type (see Figure 4a-c, h,i), differently to the smaller MNV molecule (Figure S13 for comparison). The gap states of the latter are in fact essentially localized on single Violo or NDI groups.

Thus an important conclusion from the calculations is that the electron filling of DNV pushes towards a more intermixed p-/n- type behavior and its order on the Violo and NDI groups depends on the distribution and diffusion behavior of the counter ions in the crystal.

**Electrochemical performance and optimization of DNVBr electrodes**

The capacity retention of the **DNVBr** performance in NaClO$_4$ 2.5 M was examined using variations in current and potential cut-off (Figure 5), and it was found that when using a −0.75V cut-off both the specific capacity and cyclability are comparable to that of MNV, the Coulombic efficiency being in both case ~99.7% at 4C rate. However, as expected from Figure 1c, as soon as the cut-off is set to −0.85 V a significant capacity gain is observed for **DNVBr**, irrespective of the current load. The latter is close to 85 mAh/g at 0.3 A/g (4C-rate), 79 mAh/g at 0.6 A/g (8C-rate), 76 mAh/g at 1.2 A/g (16C-rate) and 71 mAh/g at 2.4 A/g (32C-rate) while the Coulombic efficiency at 4C rate remains nearly unchanged (~99.5%). Remarkably, when applying 0.6 A/g (8C) from cycle 270 to 690 (violet curve), the capacity shifts from 70 to 65 mAh/g corresponding to a loss restricted to 0.017% per cycle.

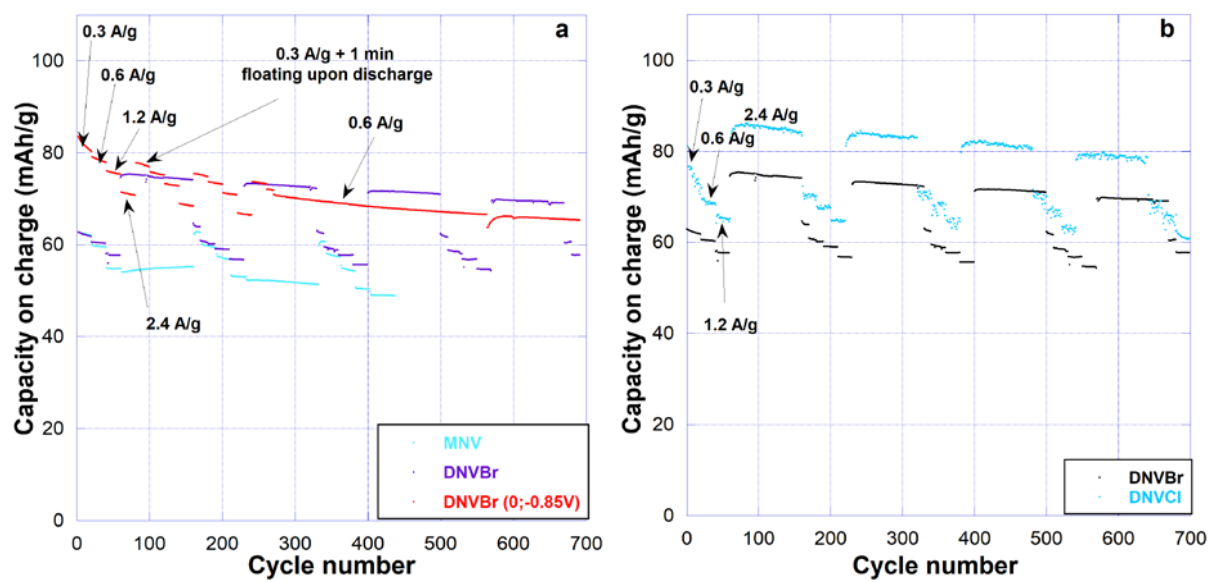

**Figure 5: (a) Capacity retentions on charge (oxidation of the material) for (violet) DNVBr and (light blue) MNV composite electrodes in NaClO$_4$ 2.5 M according to the standard cycling protocol (see experimental section). The red curve is associated with the capacity retention of DNVBr using a modified standard protocol (see experimental section) within −0.85 ≤ *E* ≤ 0 V as potential window and current loads as specified in the figure. (b) Cycling curves of DNVCl (blue) and DNVBr (black) in NaClO$_4$ 2.5 M using the standard cycling protocol (see experimental section).**

Owing to the unique chemical structure of **DNVBr** (Scheme 1), two approaches for optimizing its specific capacity can be envisioned. The first consists in turning the two terminal pyridinium groups into redox active viologen units (1 e⁻ each) by quaternization, while the second aims at decreasing the mass of the molecule by using lighter counter-anions. Taking n=3 for **DNVBr** (Scheme 1), bromine anions do represent 23% of the molecular mass. Accordingly, substituting Br⁻ for Cl⁻ and achieving the quaternization of the two terminal pyridinium moieties with methyl groups would lead to a specific capacity of 137.6 mAh/g (13 e⁻, $M_w$ = 2532.9 g/mol; considering n=3), which represents a gain of 31%. Unfortunately, the extremely low solubility of **DNVBr** in all solvents, except for trifluoroacetic acid (TFA), prevents further chemical modifications. However, the anionic exchange of Br⁻ by Cl⁻ could readily be achieved by simple immersion in an aqueous solution of NaCl 6 M at 50°C, witnessing a high anionic mobility within the DNV scaffold. Within the $-0.75 \leq E \leq 0$ V potential window, and by using 0.3 A/g (4C), this facile ionic exchange leads to a capacity gain of approx. 11.5% at 32C rate for DNVCl (Figure 5b) while the Coulombic efficiency at 4C rate is ~99.8%.

As highlighted in Figures 1 and 2, the positioning of the NDI•⁻/NDI²⁻ redox couple (III) in the vicinity of practical cut-off potentials is a means of improving the electrochemical performance of **DNVBr**. Adjusting the level of this redox couple to a higher position than that observed in NaClO$_4$ 2.5 M would indeed be one way to optimize the specific capacity. Bearing in mind that the free enthalpy of formation of each new redox state of organic materials depends on how much the charge (and/or radical) is stabilized, a convenient way of increasing the NDI-related potentials consists in augmenting the strength of the interaction between the counter ion and the NDI radical or di-anion[27]. Based on Fajans' rule[28] that $\phi=Z/r^2$, this approach is demonstrated here by changing the cation of the electrolyte from Na⁺ ($\phi_{Na^+}$ = 0.96 Å⁻²) to the much more polarizing Mg²⁺ ($\phi_{Mg^{2+}}$ = 3.86 Å⁻²).

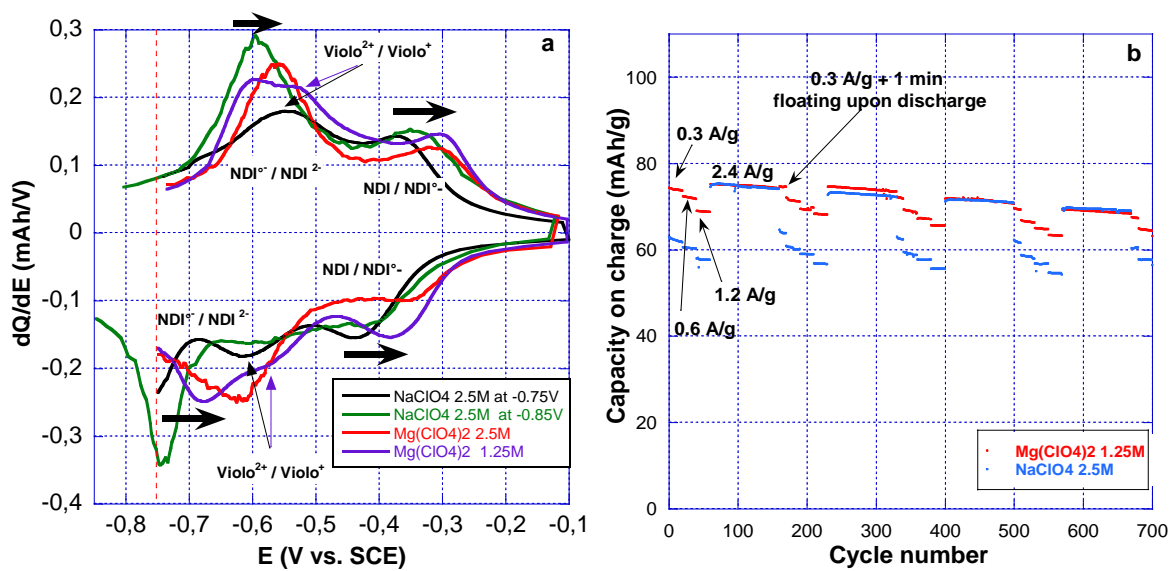

**Figure 6** a) Incremental capacity curve on cycle 4 derived from galvanostatic cycling of DNVBr composite electrodes at 0.3A/g (4C rate); b) Capacity retentions on charge (oxidation of the material) of DNVBr composite electrodes in NaClO$_4$ 2.5 M (blue) and Mg(ClO$_4$)$_2$ 1.25 M (red) using the standard cycling protocol (see experimental section).

Indeed, as pointed out by Abruña and Dichtel[29], the stabilization effect by the Mg$^{+2}$ cation induces shifts as large as +240 mV and +700 mV for NDI/NDI$^{•-}$ and NDI$^{•-}$/NDI$^{2-}$ respectively, in a porous polymer derivative when substituting 0.1 M TBAClO$_4$ with 0.1 M Mg(ClO$_4$)$_2$ in acetonitrile. In our case, substituting Mg(ClO$_4$)$_2$ for NaClO$_4$, while keeping the cation concentration constant [Na$^+$] = [Mg$^{2+}$] = 2.5 M, permits an increase in the two redox processes associated with the formation of NDI$^{•-}$ and NDI$^{2-}$ by +52 mV and +75 mV respectively (Figure 6a). Interestingly, these increments in potential are still positive (+25 mV for NDI/NDI$^{•-}$ and +27 mV for NDI$^{•-}$/NDI$^{2-}$, respectively), and remain so even with a lower concentration of salt [Mg$^{2+}$] = 1.25 M (Figure 6a). These shifts are in the same order of magnitude (100 mV) as those observed by Y. Yao in an aqueous electrolyte (estimated to be approx. 120 mV from results reported in the supplementary information[4]), and are much less pronounced than those described in the work of Abruña and Dichtel[29]. This mitigated effect in aqueous electrolytes is tentatively ascribed to the higher permittivity and solvating ability of water molecules, which presumably better shields the guest cation interaction from the **DNVBr** electron density. This hypothesis would corroborate the large amount of water involved in the insertion/de-insertion process as suggested by EQCM results. Nevertheless, Figure 6b shows that when maintaining a −0.75 V cut-off, the potential shift of the NDI$^{•−}$/NDI$^{2−}$ redox couple is large enough to induce a significant increase in the capacity, from 63 to 75 mAh/g, without having a detrimental effect on the cyclability. We note that, as expected, the magnesium electrolyte has no effect when the cut-off is set to -0.85 V during the 2.4 A/g periods (Figure 6b).

As stated before, the high NDI/Violo ratio provides substantial hydrophobicity to DNV, which results in absolute insolubility in aqueous media, as opposed to MNV[23]. Taking advantage of this property, and to further demonstrate the interest in using the DNV derivative for low-cost grid storage, long cycling experiments were conducted both in Na perchlorate and, more interestingly, in ocean water electrolytes (Figures 7a,b). In both cases, **DNVBr** shows remarkable cycling behaviors and coulombic efficiencies from 0.15 A/g (2C) (4C in NaClO$_4$) to 2.4 A/g (32C), with an impressive capacity retention of 82.8% and 77.7% upon 6500 (936 hours) and 2570 (1019 hours) cycles in NaClO$_4$ 2.5 M and ocean water, respectively, making it one of the most stable electrode materials for use as the negative electrode of aqueous batteries. We note that the pH was found to increase from neutrality to approx. 9.3 upon 6600 cycles in NaClO$_4$, which could potentially trigger a hydrolysis of the C−N bonds. However, no attempt was made to optimize the cyclability using pH buffers in the present work. Further cycling in ocean water was conducted by opening the electrochemical window from (0; −0.75V) to (0; −0.85V) and decreasing the cycling rate to 0.15 A/g (2C) and 0.3 A/g (4C). Under these conditions a gain in capacity of 20% was achieved, permitting the storage of nearly 60 mAh/g upon 3000 cycles (1800 hours). These values compare well with those of previously demonstrated aqueous batteries[4,7,8,30,31], with the added advantage of using low-cost (Na), sustainable (ocean water) and neutral electrolytes. It is worth mentioning that in ocean water the capacity drops by approx. 10 mAh/g when increasing the current load from 0.6 A/g (8C rate) to 38 mAh/g at 1.2 A/g (16C rate) whereas in NaClO$_4$ 2.5 M, under the same conditions, the drop is only approx. 3 mAh/g. Ragone tests confirm this trend (Figure S14) which matches the electrolyte conductivity values (Table S2). Additionally, because similar Ragone profiles are obtained for both NaCl and NaClO$_4$ with the same concentration (2.5 M, Figure S14), it is the ionic conductivity of the electrolyte rather than the nature of the anion that rules the kinetics of these electrochemical reactions.

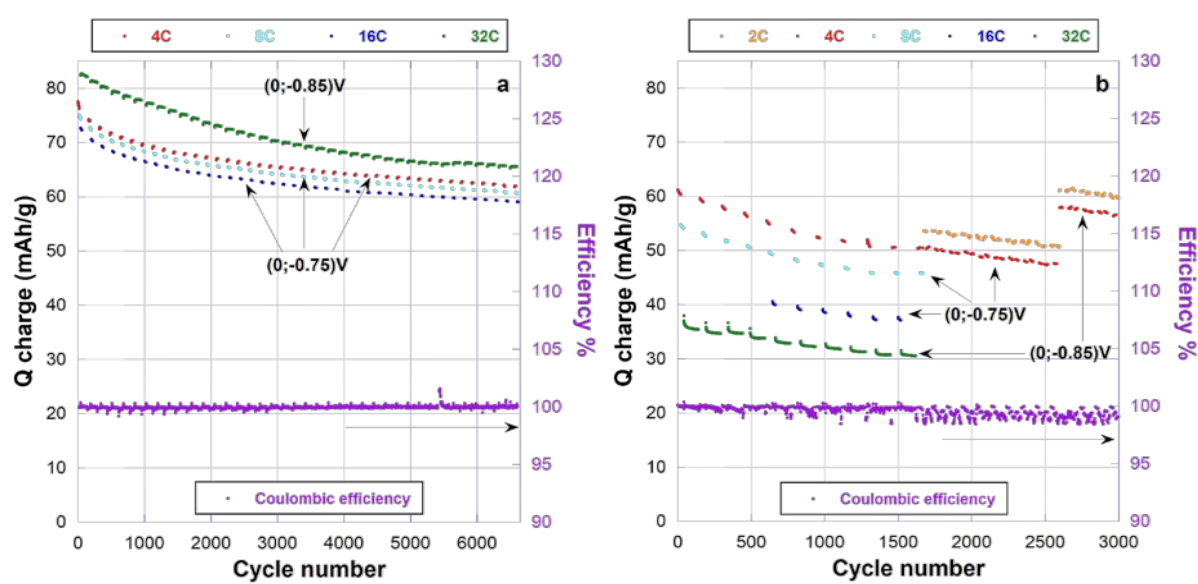

**Figure 7:** Capacity retention on charge (oxidation of the material) and corresponding Coulombic efficiency curves for DNVBr composite electrodes in (a) NaClO$_4$ 2.5 M and (b) ocean water.

In conclusion, by comparison to conventional inorganic and organic materials previously reported, several unique advantages can be gleaned from the specific molecular structure of DNV. Indeed, *(i)* NDI brings both a high hydrophobicity which mitigates dissolution issues during battery operation and a large specific capacity while *(ii)* the propyl linker provides flexibility as well as robustness, which prevents the capacity fading observed for NDI[23], and *(iii)* the viologen units are in the "charged state" and allow p-type redox processes (anionic), which could potentially mitigate both the ionic depletion occurring within the porosity of ultra-thick electrodes (> 1 mm)[32,33] as well as electrode volume variations upon cycling.

Aside from some rare examples[34,35], the use of organic materials for electrochemical storage in both aqueous and non-aqueous media is hampered by the large quantities of carbon additives (typically >30 wt.%)[36] that are required in order to compensate for their relatively low electronic conductivity. Added to the intrinsically low volumetric density of organic materials, this flaw makes the efficient design of the electron percolating[37] network and the optimization of the electrode thickness[5,38,39] even more critical to reducing the cost of stored energy. These two key points, scarcely studied in the literature[5,40], were investigated for **DNVBr** composite electrodes by using varying amounts of carbons, mixing protocols and areal capacity. We note that carbon nanotubes were not considered in this work owing to their elevated price, even though they have been shown to significantly reduce the percolation threshold of electroactive organic-based composite electrodes[5,40]. Contrary to most of the electroactive organic materials published thus far[1], we demonstrate below that at 1C, a rate that is compatible with renewable energy storage, the best specific capacity per mass of the whole electrode is obtained with a carbon black content as low as 10wt%.

Let us first consider the electrical property of a standard thin electrode of 0.7 mAh/cm$^2$ where 25% carbon additive was hand-mixed with **DNVBr** (referred to as **DNVBr**-HM-25%). In this case, EIS measurements show that the resistance associated with the semi-circle decreases slightly from approx. 5.5 ohms to 3.5 ohms during the reduction process from the initial state to −0.64 V, and then remains constant irrespective of the state-of-charge (Figure S15b). Given the high conductivity of the electrolyte (Table S2), this behavior indicates that **DNVBr** charge transfer resistance is not detectable, and that it is instead the response of the percolating carbon additive network which is measured. Such low cell resistance is expected to induce a barely detectable polarization of 4.5 mV if the current increases from 0.3 A/g (4C rate) to 2.4 A/g (32 C rate). This point is illustrated in Figure S16 where a negligible polarization and a drop in specific capacity are indeed observed while the current load is multiplied by a factor of 8 both on discharge and on charge, from 4C to 32C. There are thus two points worth noting from these results: (i) 25% carbon additive constitutes a large excess of carbon additive and (ii) **DNVBr** seems to be associated with extremely fast intrinsic kinetics (that includes charge transfer, phase boundary displacement, solid state ionic diffusivity and electron conductivity), which do not appear to substantially limit the electrode properties, even at rates of up to 32C.

The electrical behavior of **DNVBr** itself was therefore unraveled by decreasing the weight fraction of the carbon additive to 10 wt% (**DNVBr**-HM-10%). The resulting EIS measurements do indeed show a correlation between the state-of-charge and the value of the resistance that is ascribed to a charge transfer mechanism (Rct) (Figure 8a, fits with the data reported in Figure S17): (i) a marked decrease in Rct during the reduction process, (ii) an increase in Rct upon completion of the reduction, and (iii) a reverse trend during oxidation. We note the presence of the same evolution during the second cycle (Figure S17). The intriguing increase of Rct observed at the end of the reduction process can tentatively be ascribed to the drastic change in the electronic state of the molecule from a dual-radical form (NDI$^{•-}$ and Violo$^{•+}$) to a radical/quinoid one (NDI$^{2-}$ and Violo$^{•+}$). However, further research needs to be devoted to this point since this "electronic" transformation is paired with the structural evolution of the solid (P1 to P2), as shown in Figure 3. The results nonetheless show that the reduction of **DNVBr** is accompanied by a significant decrease in Rct, which supports DFT findings. They also show that, except for the most oxidized states, Rct remains as low as 13 to 45 ohms, which opens up the possibility for a further increase in active material loading. In order to confirm this, we tested **DNVBr** without any carbon additive. We found only one other example of an organic material having been shown to be electro-active without carbon additive[34]. Results reported in Figure S6 demonstrate the most unexpected capability of **DNVBr**, showing it to deliver nearly 50 mAh/g$_{electrode}$ at C/5 rate and 40 mAh/g$_{electrode}$ at C-rate.

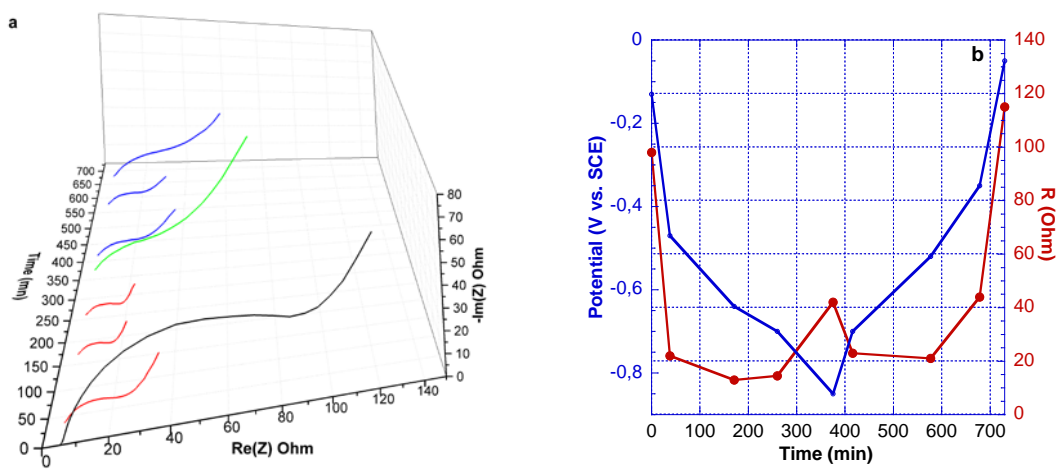

**Figure 8 a) EIS spectra of a DNVBr electrode prepared by hand-mixing with 10 wt.% carbon additive during the first cycle; b) evolution of the charge transfer resistance (Rct) during the first cycle as derived from the fit of the semicircle in a), along with potentials at which the EIS measurement was performed. EIS spectra and corresponding fits are gathered in Figure S17.**

The effect of the surrounding composite electrode on the overall performance was further investigated by studying the influence of a ball milling step (BM) using different carbon contents. These samples are referred to as **DNVBr**-BM-X%, where X stands for the weight fraction of carbon additive (5 < X < 20). SEM images, as well as specific surface values derived from BET analysis, are reported in Figures S18, S19 and S20, confirming that the ball milling step strongly favors inter-particle contacts. The cyclability of these electrodes is reported in Figure S21. Fits to the EIS data (Figure S22 and S23) clearly show that as the carbon content drops from 20 wt.% to 5 wt.% the resistance (Rct) sharply increases from 11 to 1850 ohms (Figure S23). This trend is most prominent between 10 wt.% and 5 wt.%, suggesting that a threshold lies in between these two compositions. Ragone plots performed upon oxidation (Figure 9a) indeed show that with 5 wt.% carbon additive, 70% of the full capacity (which is 94 mAh/g$_{DNVBr}$ as observed for **DNVBr**-20-BM) is obtained at 0.22C, while with 10 wt.% of carbon additive, full capacity can be attained. Importantly, out of the four samples **DNVBr**-10-BM shows the best specific capacity up to C rate in terms of electrode mass (Figure 9b), with a gain as high as 30% at C/4 rates (82 mAh/g$_{electrode}$) vs. **DNVBr**-20-BM (69 mAh/g$_{electrode}$). Increasing the amount of carbon additive above 10wt% is therefore only necessary for high-rate uses. In this regard, it is to be noted that **DNVBr**-BM-15 and **DNVBr**-BM-20 can still achieve 47 mAh/g and 64 mAh/g per mass of electrode, respectively, at a currents as high as 9.6 A/g (88 C), which competes well with best results obtained for similar areal capacity organic-based electrodes as reported by Oyaizu and Nishide[5] (0.8mAh/cm$^2$ PTAm-SWNT hybride electrodes with optimized current collector/electrode interface), and by Y. Yao[4] (0.5 mAh/cm$^2$ electrodes of PPTO). It is also instructive to compare these results to the state-of-the-art of inorganic negative electrode materials for Aq batteries which are Na(Li)Ti$_2$(PO$_4$)$_3$ (referred to as NTP and LTP respectively). The relatively high density of these materials (approx. 2.6 g/cm$^3$ [41]) leads to theoretical volumetric capacity of approx. 260-310 mAh/L. In comparison **DNVBr** shows approx. 170 mAh/L using a density of 1.6 g/cm$^3$ [4]. However, the carbon content of LTP and NTP based electrodes is usually high (approx. 6-15wt% of carbon coating and up to 30-40% of carbon additive[42]. Accordingly, it is not sure the high density of the active inorganic materials (NTP or LTP) is much of an advantage in terms of final volumetric energy density compared to that inferred by **DNVBr** which shows a maximum gravimetric capacity with only 10 wt% of carbon additive at C-rate. In addition, dissolution of NTP and LTP is also of great concern in water or molar range electrolytes with the formation of insulating phosphate layer at the surface of the particles and/or Ti-complexes with anions of the electrolyte within the porosity of the electrode[43].

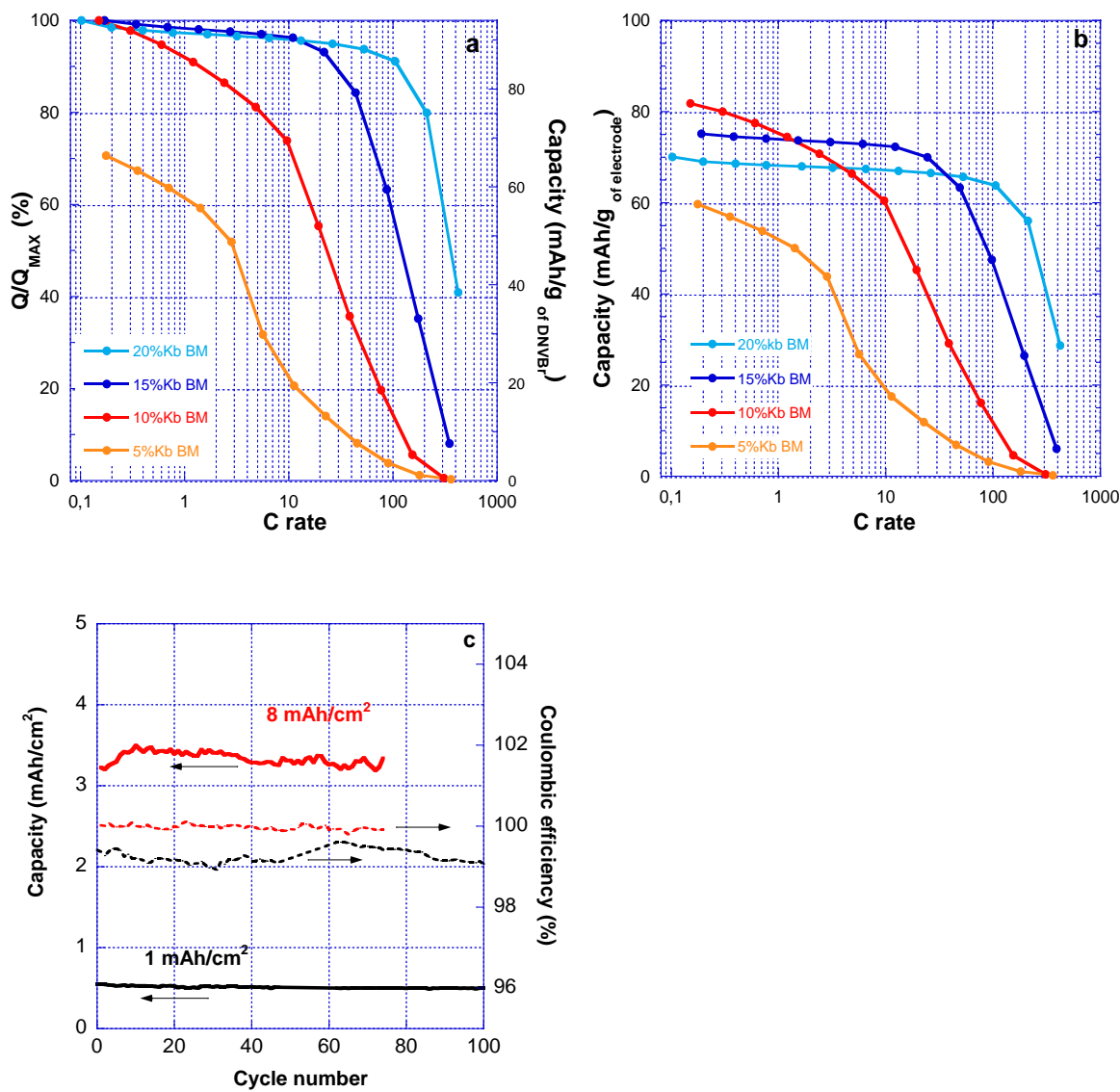

**Figure 9: a)** Ragone plots of DNVBr performed during oxidation with different percentages of carbon: 20 wt.% Kb (light blue), 15 wt.% Kb (dark blue), 10 wt.% Kb (red), 5 wt.% Kb (orange). The relative capacity "Q/Q$_{Max}$" in a) refers to the maximum one Q$_{max}$, as obtained for 20 wt.% Kb BM at C/10 rate and the specific capacity refers to the mass of active material, in b) the specific capacity refers to the mass of electrode. **c)** Comparison of experimental surface capacity, cyclability and Coulombic efficiency for 1 and 8 mAh/cm² electrodes of DNVBr-HM-25 in saturated NaClO$_4$ electrolyte at 1C rate.

Due to the fact that aqueous battery technology requires electrodes to be several mAh/cm² so as to reach the 100 $/kWh target, we also demonstrate in Figure 9c that the performance of **DNVBr**-HM-25 implemented in a thick electrode of 8 mAh/cm² (76.5 mg/cm² of **DNVBr**, 0.53< thickness < 0.78 mm, Figure S24) shows similar cyclability and Coulombic efficiency at 1C rate to a 1 mAh/cm² (9.6 mg/cm² of **DNVBr**, 65 μm) (Figure 9c), yet possesses a much higher areal capacity (3.4 mAh/cm² vs. 0.55 mAh/cm²).

In conclusion, **DNVBr** is a unique and highly promising electroactive organic material that can simultaneously exchange both inexpensive as well as abundant cations (such as Na and Mg) and anions (Cl⁻), and can reach more than 100 mAh/g at an optimal potential of −0.55 V vs. SCE with remarkable capacity retention and Coulombic efficiency in several neutral electrolytes, including ocean water. Thanks to the intrinsically high kinetics of the material, straightforward electrode engineering was shown to enable a significant optimization of the overall performance by increasing the active material content from 75 to 95%, and the surface capacity from 0.65 to 3.6 mAh/cm$^2$.

**Electrochemical performance of full cells**

To further demonstrate the interest of **DNVBr**, a full organic cell was assembled in the discharge state using a commercial p-type TEMPO derivative (4-HydroxyTEMPObenzoate, referred to as 4HT) as the active material of the positive electrode. A comprehensive study of the 4HT material being out of the scope of this paper, we put forward a proof of concept of the cell using three electrodes set up in order to follow the electrochemical response of the two electrodes. We note that, at this stage of the study, and with 4HT being slightly soluble in NaClO$_4$ 2.5 M, we used a more concentrated NaClO$_4$ electrolyte (arbitrarily saturated). It is worth noting this cell is a hybride, p-type (anionic) rocking chair battery, and dual ion since **DNVBr** still needs to draw sodium cations from the electrolyte. To mimic the galvanostatic cycling of a real battery, no potential restrictions were applied to the two working electrodes. Rather, the cell voltage was controlled between 0 and 1.8 V. Under these conditions, the best capacity ratio was found to be nearly 1.0, resulting in the highest average discharge voltage of 1.07 V (Figure 10a) and the best cyclability (Figure 10b). This output average voltage is comparable to the currently most promising Na (1.1-1.4 V), Li aqueous based systems (1.1 V)[4,44] and mixed Li/Na cells[6]. The specific energy density of this new cell is 36 Wh/kg by mass of the two active materials at C rate, which is competitive with recent sodium-based aqueous batteries ($27^7$-$42^{6,44}$ Wh/kg) but still lower than the Li-aqueous mixed inorganic/organic battery recently achieved (92 Wh/kg[7]). We note, however, that all these batteries use at least one inorganic material which is associated with larger grey energy[45]. Moreover, because of the high ionic conductivity of the electrolyte (9.6 S/cm) and rapid reaction of the electrodes in the present cell, 98.4% of the maximum discharge capacity was retained at 8C rate, while at nominal rates of 16C (1.2 A/g) and 32C (2.4 A/g), 97.1% and 95% of the capacity was still available, respectively. When compared to the state-of the art of Na and Li aqueous batteries, an excellent cyclability could be achieved, with 80% of the initial capacity being retained after 1200 cycles (600 hours). The Coulombic efficiency stays remarkably high (>99.93%) at high rates, and higher than 99.57% at 1C rate (Figure 10b). With respect to corrosion issues, it is instructive to note that charging the cell up to 1.8V did not provoke any strong deviation of the pH which self-buffers to around 5 (Figure 10b).

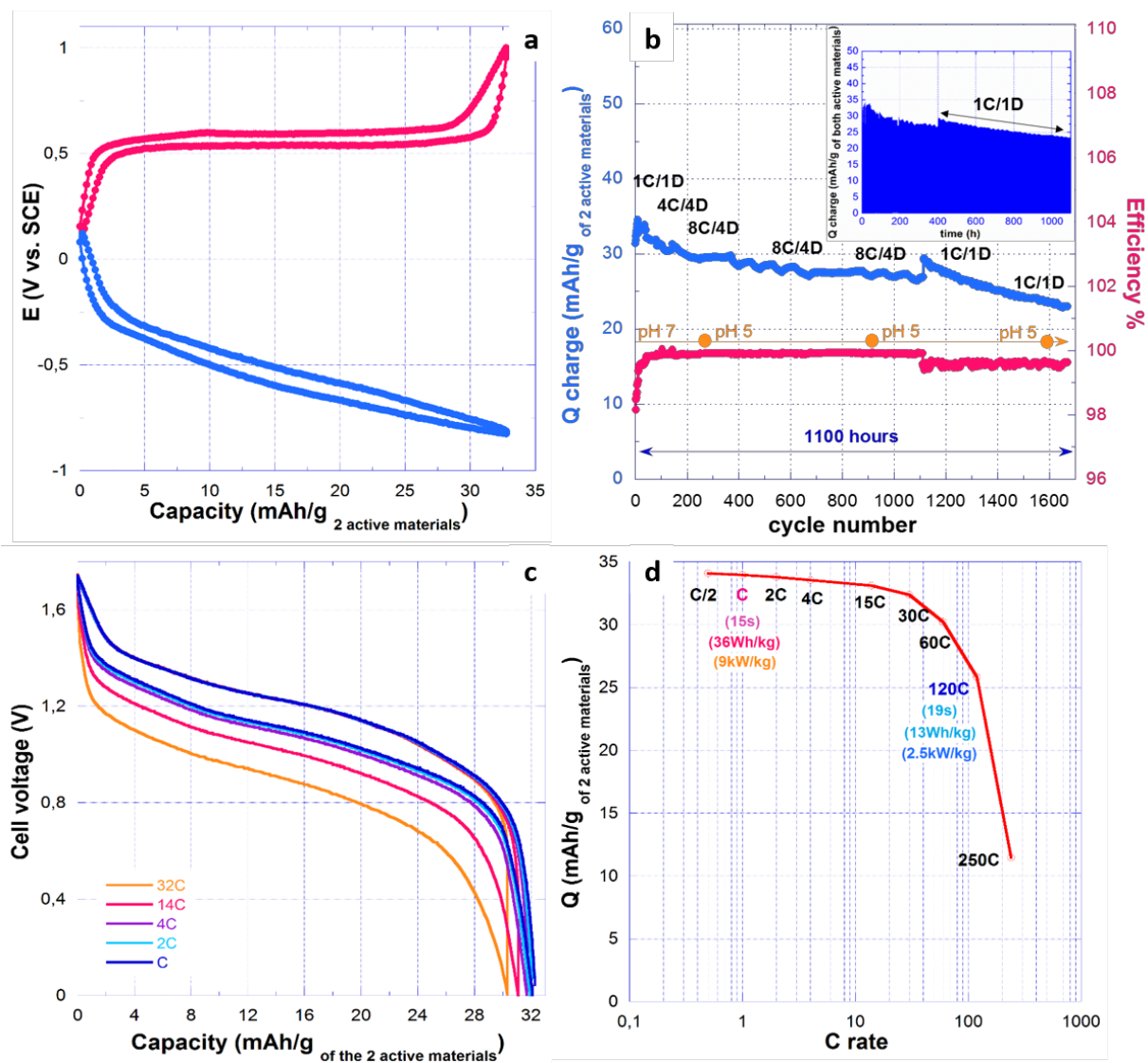

Figure 10: a) Potential-capacity profile of both 4HT and DNVBr electrodes of 0.65 mAh/cm$^2$, b) Cyclability of the full cell at different C rates along with pH values of the electrolyte. C and D stand for charge and discharge rate, respectively. The inset shows that most of the cycling time was performed at C-rate, c) Voltage-capacity profile of the full cell, at C rate on charge, and from C to 32C rate on discharge, and d) the corresponding Ragone plot.

**Conclusion:**

All of the above results serve to demonstrate the uniqueness and efficiency of the proposed di-block bipyridinium-naphthalene diimide oligomer (DNV) in maintaining a high capacity, as well as in sustaining long-term cycling, in several molar range and neutral aqueous electrolytes, such as ocean water, for several thousand cycles. Furthermore, the capacity for anionic exchange, electrolyte tailoring and the design of highly-loaded (8 mAh/cm$^2$) or devoid of conducting additives electrodes is shown to provide excellent prospects for future performance enhancement. Coupled with a commercial TEMPO derivative and a sodium electrolyte, this full organic aqueous battery ensures a long cycle life, extremely fast kinetics, as well as promising energy density values, thereby placing **DNVBr** among the top contenders for large-scale energy storage.

**Acronyms**
EQCM, electrochemical quartz crystal microbalance; EIS, electrochemical impedance spectroscopy; NDI, Naphthalene diimide; Violo, Bipyridinium salt; XRD, X-Ray diffraction

**Acknowledgements**
JG is thankfull to Prof. Yan Yao (Houston University) and Prof. Michel Armand (CIC Energigune) for fruitful discussions. JG and SP wish to thank: Dr. Antonio Jesus Fernandez Ropero (IMN), Dr. Guillaume Ledain (IMN), and Dr. Nicolas Stephan (IMN) for their help with SEM pictures; Dr. Christelle Gautier (MOLTECH Anjou) for her technical assistance with EQCM measurements performed at Moltech Anjou in Dr. E. Levillain's group; Dr. V. Silvestre and B. Charrier (CEISAM, University of Nantes) for their help with NMR spectrometers, and Prof. Bernard Humbert (IMN) for fruitful discussions regarding FTIR data.
JG is grateful to synchrotron SOLEIL for funding proposal 20160193 on the CRISTAL beamline, and Dr. Erik Elkaim (SOLEIL) for technical assistance with the beamline.
JG is thankful to Nathalie Cochennec-Laureau (IFREMER) for providing salinity measurements and Dittmar's method for ion composition determination.
All of the above-mentioned authors contributed to the present manuscript, and all have approved the final version.

## Experimental Section

*Characterization techniques*:

$^1$H-NMR spectra were acquired using a Bruker ARX 300 MHz spectrometer. $^{13}$C-NMR spectra were recorded on a Bruker 500 MHz, operating at 125MHz, using a dual 1H/13C cryoprobe. Spectra were recorded at room temperature. Chemical shifts are reported in ppm, and coupling constant in Hz. Multiplicity is presented in the following way: s = singlet, d= doublet, t = triplet, q = quintuplet, m = multiplet. Thermal analyses were performed with a NETZSCH STA 449F3 device under Argon atmosphere. Fourier transform infra-red spectroscopy (FTIR) spectra were collected with a Bruker Vertex 70 device in ATR mode, using a DTGS detector at a resolution of 4 cm$^{-1}$. Scanning electron microscopy (SEM) was performed using a JEOL JSM-7600F microscope. Energy-dispersive X-ray spectroscopy (EDS) was performed with a Hitachi HF-2000.

**Characterization techniques**

UV-Vis Spectroelectrochemistry of **DNVBr** was conducted using a BUNDLE-FLAME-ABS Ocean Optics spectrometer under N$_2$ atmosphere in an airtight glove bag. Each scan (0.7s/scan) is the average of 50 fast-scans of 1.4 10$^{-2}$ s recorded from 250 to 1050 nm. Electrochemical quartz-crystal microbalance (EQCM) was performed with a Stanford Research Systems QCM200 Digital Controller and a QCM25 Crystal Oscillator (5 MHz Crystal). O*perando* X-ray diffraction was performed at SOLEIL synchrotron on the CRISTAL beamline using a 2D MAR detector. We used a home-made pouch cell bearing three electrodes sealed in Mitsubishi Escal Neo transparent films. The working electrode containing 70 wt.% active material, 25 wt.% Ketjen Black EC-600JD, and 5 wt.% PTFE was precisely aligned in the center of the pouch cell; a carbon pseudo-reference electrode located on one side, and a three-fold oversized carbon counter electrode (80 wt% activated carbon Norit 1600, 10 wt% Ketjen Black, and 10 wt% PTFE) on the other side. All the electrodes were prepared following the procedure described in the next section ("Electrode preparation") and all the pouch cells were mounted in a glove bag under constant argon flow. Electrodes were insulated from one another using a microfiber glass separator impregnated with 150 μL of electrolyte before sealing at 180°C. The reduction-oxidation of the electrode was performed using a potential cycling with galvanostatic acceleration (PCGA) technique between (0; -0.88V) vs. SCE. The duration of each potential step of 20 mV was imposed either by a limiting current (corresponding to C/5) or by its maximum duration set to 2.5 minutes. To deepen the reduction process, the duration of the two last steps at -0.86 and -0.88V were extended to 5 minutes.

Electrochemical tests were performed using either SP 300 or VMP potentiostats from Bio-logic SAS (Claix, France). **DNVBr** was characterized by cyclic voltammetry, potentiodynamic and galvanostatic cycling in $NaClO_4$ (1.25M and 2.5 M), ocean water (coast of Pornic, France, salinity of S=34) and in $Mg(ClO_4)_2$ (1.25 M and 2.5M) aqueous electrolytes. Based on Dittmar's Law ($C_{ion}$(mol/kg)=S*(Ionic factor)) was used to determine the precise composition of the ocean water as follows:

| Ions | Ionic factor (mmol/kg) Dittmar Law | C (mol/kg) | C (Mol/l) |
|---|---|---|---|
| $Na^+$ | 13.402 | 0.455668 | 0.46683187 |
| $Mg^{2+}$ | 1.509 | 0.051306 | 0.052563 |
| $Ca^{2+}$ | 0.2938 | 0.0099892 | 0.01023394 |
| $K^+$ | 0.2916 | 0.0099144 | 0.0101573 |
| | | | |
| $Cl^-$ | 15.597 | 0.530298 | 0.5432903 |
| $SO_4^{2-}$ | 0.807 | 0.027438 | 0.02811023 |
| $Br^-$ | 0.02404 | 0.00081736 | 0.00083739 |
| $F^-$ | 0.00195 | 0.0000663 | 6.7924E-05 |

DNVCl was characterized by galvanostatic cycling in $NaClO_4$ 2.5 M. 4HT was characterized by galvanostatic cycling insaturated$NaClO_4$ aqueous electrolytes. A standard cycling test is defined as a series of successive galvanostatic periods of 20 cycles at 0.3A/g (4C), 20 cycles at 0.6A/g (8C), and 20 cycles at 1.2A/g (16C) in a (0; -0.75V) potential window, followed by 100 cycles at 2.4A/g (32C) in a (0; -0.85V) potential window. A threefold-size counter-electrode (80 wt% activated carbon Norit 1600, 10 wt% Ketjen Black, and 10 wt% PTFE) was used to ensure that its potential remained in the 0; +0.3 region vs. the Saturated Calomel Electrode (SCE), which was used as the reference electrode. All tests were performed within either a glove bag (Aldrich) under $N_2$ flow or a glove box (Braun) under Ar with less than 1ppm $O_2$. When glove bags were used, electrolytes were degassed with argon bubbling for 1h within the glove bag prior to use. Experimental errors related to capacity values were always below 5% using our protocols and set-up, and both glove bags and glove box environments were found to lead to similar electrochemical results.

Electrochemical impedance measurements were performed from 180 kHz to 100 mHz with a perturbation voltage of 7mV. Power tests on charge were carried out using the "single charge technique" consisting of OCV relaxations until $\Delta U/t = 1$ mV/h between each current pulse from $I_{max}$ to $I_{min}$. All experiments were conducted twice in order to ensure reproducibility.

*Synthetic Procedures*:

All commercially obtained solvents and reagents were used without further purification unless noted below. The 1,4,5,8-naphthalenetetracarboxylic dianhydride, 4,4'-bipyridyl, 3-bromopropylammonium bromide and sodium perchlorate were purchased from Sigma-Aldrich.

Compound **3**: A mixture of naphthalene tetracarboxylic dianhydride (536 mg, 1.99 mmol), 3-3-bromopropylammonium bromide (1.64 g, 7.5 mmol) and $Et_3N$ (1.0 mL) in AcOH (10 mL) was refluxed for 24 h. The reaction mixture was filtered and thoroughly washed with water and MeOH to give **3**, as a white solid (900 mg, 89%). This compound was used in the next reaction without further purification. $^1$H NMR (300 MHz, DMSO-d$_6$): δ 8.69 (s, 4H), 4.19 (t, *J* = 6.9 Hz, 4H), 3.64 (t, *J* = 6.9 Hz, 4H), 2.24 (q, 4H).

Compound **DNVBr**: In a sealed tube, N,N'-Bis(2-bromoethyl)-1,4,5,8-naphthalenetetracarboxylic 1,8:4,5-diimide **3** (500 mg, 0.984 mmol) was added to 4,4'-bipyridine (154 mg, 0.984 mmol) (**4**) in anhydrous DMF (10ml). The reaction mixture was reacted for 4 days at 130°C. After cooling down to room temperature, the dark brown precipitate was filtered and thoroughly washed with methylene chloride yielding 180 mg of final product **5**. The dark brown product obtained is totally insoluble in organic solvents, preventing its characterization by Size Exclusion Chromatography and by Mass spectrometry analyses, and therefore precluding the determination of its polydispersity. Accordingly, the yield of the last step could not be accurately determined. $^1$H NMR (300 MHz, TFA-d$_1$): δ 9.34 (m, 13H), 9.23 (d, J = 6.4 Hz, 4H), 8.99 (m, 18H), 8.64 (m, 17H), 5.13 (m, 14H), 4.60 (m, 18H), 2.82 (m, 14H). FT-IR/ATR (cm$^{-1}$):3382, 2990, 1702 (C=O imide asym.), 1656 (C=O imide sym.), 1640 (C=N+), 1577(C=N), 1449, 1376, 1331, 1242, 1178, 1050, 1013, 970, 814, 766.

Ionic exchange (**DNVCl**): 50 mg of **DNVBr** was dispersed in aqueous 6 M solution of NaCl. The suspension was stirred for 4 days at 50°C. Subsequently, the precipitate was washed several times with water, dried at 60°C under vacuum overnight to finally obtain 40 mg of DNVCl. The final product was characterized by SEM-EDS (Figure S25), $^1$HNMR (Figure S26) and TGA-MS/DSC (Figure S27). $^1$H NMR (300 MHz, TFA-d$_1$): δ 9.36 (m, 16H), 9.23 (d, J = 6.4 Hz, 4H), 8.99 (m, 22H), 8.64 (m, 20H), 5.12 (m, 16H), 4.59 (m, 17H), 2.82 (m, 16H).

*Electrode preparation*: All electrodes contained 5wt% PTFE as the binder and only the **DNVBr**/carbon additive ratio was varied. However, unless otherwise specified, the electrode composition remained 70 wt% active material, 25 wt% Ketjen Black, and 5 wt% PTFE. The **DNVBr** BM electrodes were prepared by ball milling the mixture of active material and conductive carbon for one hour at 700 rpm using the Pulverisette 7 classic line (Fritsch). To this end, a total amount of 100 mg of the mixture was introduced along with 0.25 mL of water and three silicon nitride balls. The slurry was then dried and mixed with PTFE in order to obtain electrode compositions of **DNVBr**/KB/PTFE ranging from 75/20/5 to 90/5/5. The mixture was pressed at 5 tons on a stainless steel (AISI 304L) grid current collector. Unless otherwise specified, the electrodes were in the 0.6 to 0.7 mAh/cm$^2$ range.

Thick electrodes of 8 mAh/cm² were made in two steps. Firstly, a pellet was obtained by pressing a mixture of 70 wt% active material, 25 wt% Ketjen Black, and 5 wt% PTFE at 5 tons. Secondly, this pellet was placed between two stainless steel grids and pressed at 2 tons in order to mitigate contact resistance with the current collector.

*Density Functional Calculations:* Density functional calculations were performed using the AIMPRO code [46–48] with the LDA-PW92 exchange-correlation functional.[49] Periodic boundary conditions were applied to very large orthorhombic unit cells, 220×50×50 au, and energies converged to <10$^{-7}$ Ha. Electronic levels were filled using a Fermi occupation function to aid convergence, with kT=0.04eV. Charge density was fitted using plane-waves with an energy cut-off of 220 Ha. Charged systems were compensated in the conventional manner via a uniform jellium background charge. Relativistic pseudopotentials, generated using the Hartwigsen-Goedecker-Hutter scheme[50], were expanded via Gaussian-based polynomials up to *l*=2, with 38/40/40/12 independent Gaussian functions per carbon/nitrogen/oxygen/hydrogen. Atomic charge states were calculated using Mulliken population analysis. For the neutral species we also checked alternative spin states, confirming the spin restricted to be the most stable.

# Supporting information

**Intermixed Cation-Anion Aqueous Battery Based on Extremely Fast and Long Cycling Di-Block Bipyridinium-Naphthalene Diimide Oligomer**

Sofia Perticarari,[a,b] Tom Doizy,[a] Patrick Soudan,[a] Chris Ewels,[a] Camille Latouche,[a] Dominique Guyomard,[a] Fabrice Odobel,[b,*] Philippe Poizot,[a,*] and Joël Gaubicher[a,*]

a. Institut des Matériaux Jean Rouxel (IMN), Université de Nantes, UMR CNRS 6502, 2 rue de la Houssinière, B.P. 32229, 44322 Nantes Cedex 3, France
E-mail: Joel.Gaubicher@cnrs-imn.fr and Philippe.Poizot@cnrs-imn.fr

b. Chimie et Interdisciplinarité, Synthèse, Analyse, Modélisation (CEISAM), UMR CNRS 6230, Université de Nantes, 2 rue de la Houssinière, B.P. 92208, 44322 Nantes Cedex 3, France
E-mail: Fabrice.Odobel@univ-nantes.fr

- **Preparation of 1-propyl-4-pyridyl-pyridinium bromide**

In sealed tube 4.7 mL (51.2 mmol) of 3-bromo-propane and 400 mg (2.56 mmol) of 4,4'-bipyridine were heated overnight at 80°C. The reaction mixture was cooled to room temperature and 20 mL of diethyl ether are added. The resulting yellow precipitate was filtered, abundantly washed with diethyl ether and finally dried under vacuum affording 495 mg of 1-propyl-4-pyridyl-pyridinium bromide as a yellow solid (yield: 70%). The $^1$H NMR is in agreement with that previously reported. $^1$H NMR ($\delta$ [ppm], TFA-$d_1$): 1.16 (t, J = 7.4 Hz, 3H), 2.23 (t, J =7.4 Hz, 2H), 4.80 (t, J =7.4 Hz, 2H), 8.65 (m, 4H), 9.13 (d, J = 6.8 Hz, 2H), 9.20 (d, J = 7.1 Hz, 2H).

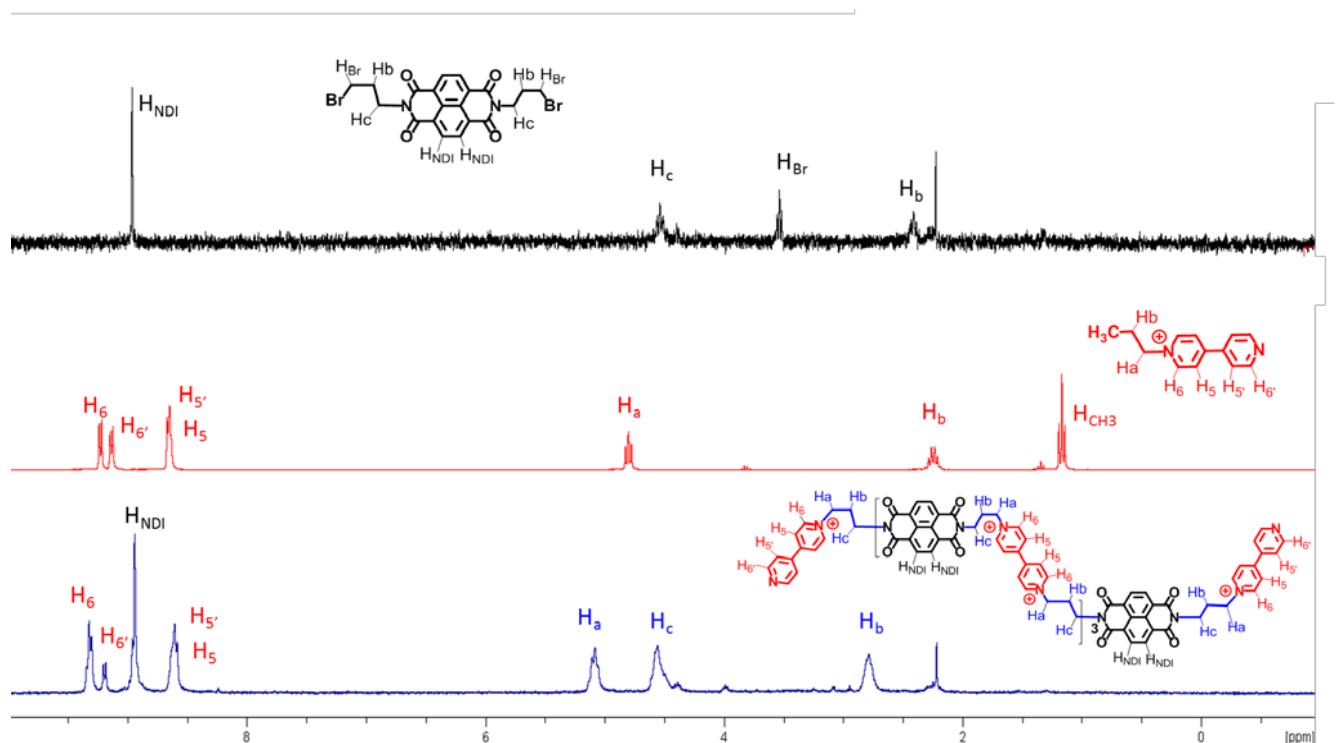

***Figure S1:*** $^1$H NMR spectra of DNVBr (blue) along with those of 1-propyl-4-pyridyl-pyridinium bromide (red) and NDI 3 (black) recorded in TFA-$d_1$. Due to the symmetry of the molecule, only half of the protons are labelled, knowing that the others are magnetically equivalent. For example, $H_6$ in bipyridines are similar to $H_2$ and so forth.

Note that the attribution of the signals in the $^1$H NMR spectrum of **DNVBr** was made by comparison with the spectrum of each component, which are NDI **3** and 1-propyl-4-pyridyl-pyridinium bromide (**Figure S1**). The most deshielded signal in the aromatic region is assigned to the $H_6$ protons of the alkylated pyridine, while that just below is due to the $H_{6'}$ protons of terminal non alkylated pyridine. The proton of naphthyl unit of NDI appear as a singlet at 8.99 ppm well separated from the other peaks, preventing any confusion. The aromatic protons $H_5$ and $H_{5'}$ in the bipyridine are almost equivalent and show up as a multiplet, in agreement with those in 1-propyl-4-pyridyl-pyridinium bromide. In the aliphatic region, the assignment is relatively straightforward, because $H_c$ (CH$_2$ next to bisimide in NDI) of **DNVBr** was not shifted compare to its position in NDI **3** (**Figure S1**). $H_a$ (CH$_2$ next to viologen) is the most deshielded aliphatic signal, while $H_b$ is naturally the most shielded as it is not directly adjacent to an electron withdrawing substituent.

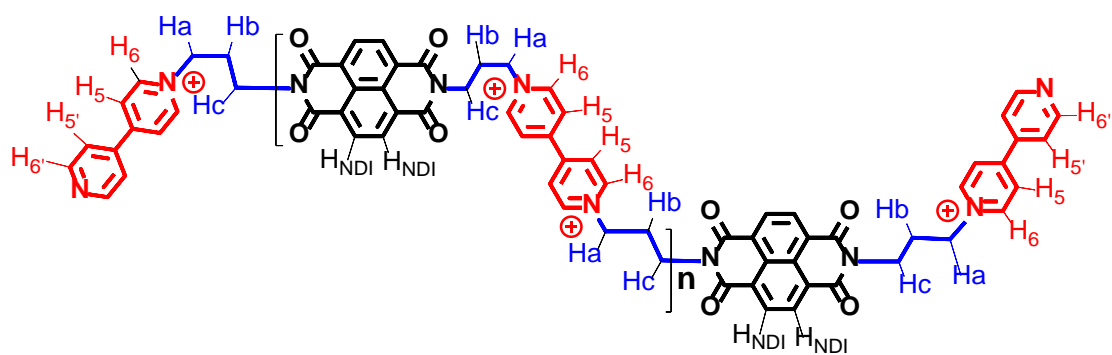

**Scheme S1.** *General structure of the **DNVBr** oligomer.*

The general structure of the oligomer (**Scheme S1**) enables to deduce the number of protons for each peak as a function of the number of motifs "n" in the oligomer and is exemplified below:

$H_{6'}$: 4 protons
$H_5$ + $H_{5'}$ : 4n + 8 protons
$H_6$: 4n + 4 protons
$H_{NDI}$: 4n + 4 protons
$H_a$ = $H_b$ = $H_c$ = 4n + 4 protons

Accordingly, the degree of oligomerization of **DNVBr** can be deduced by the integration of specific signals in its $^1$H-NMR spectrum. Particularly, the ratio between the $H_a$ protons (4n+4) and $H_{6'}$ protons (4) is equal to n+1 or the ratio between $H_5$+ $H_{5'}$ protons (4n+8) and $H_{6'}$ protons (4) is equal to n+2. The average experimental value being 3, thus n = 3±1.

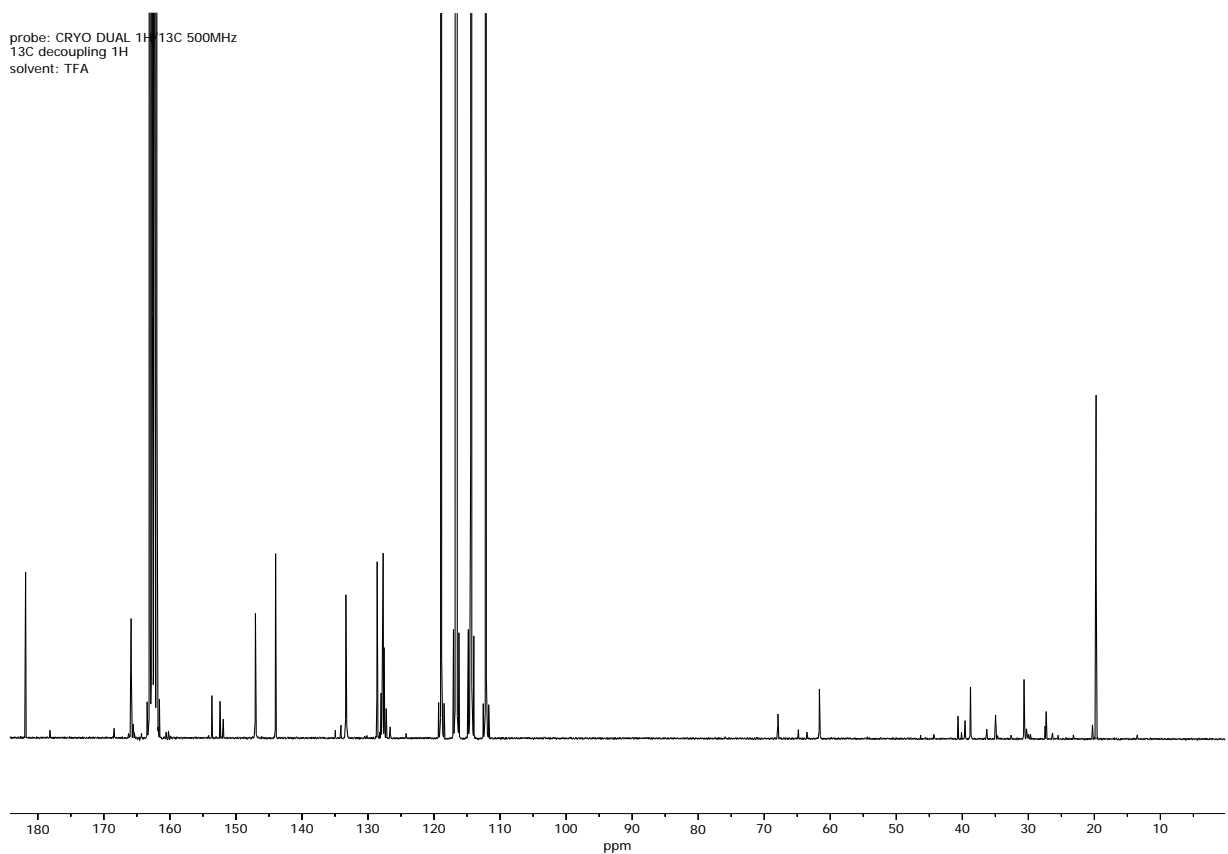

***Figure S2.*** *500 MHz $^{13}$C NMR spectrum of DNVBr recorded in TFA-d$^1$.*

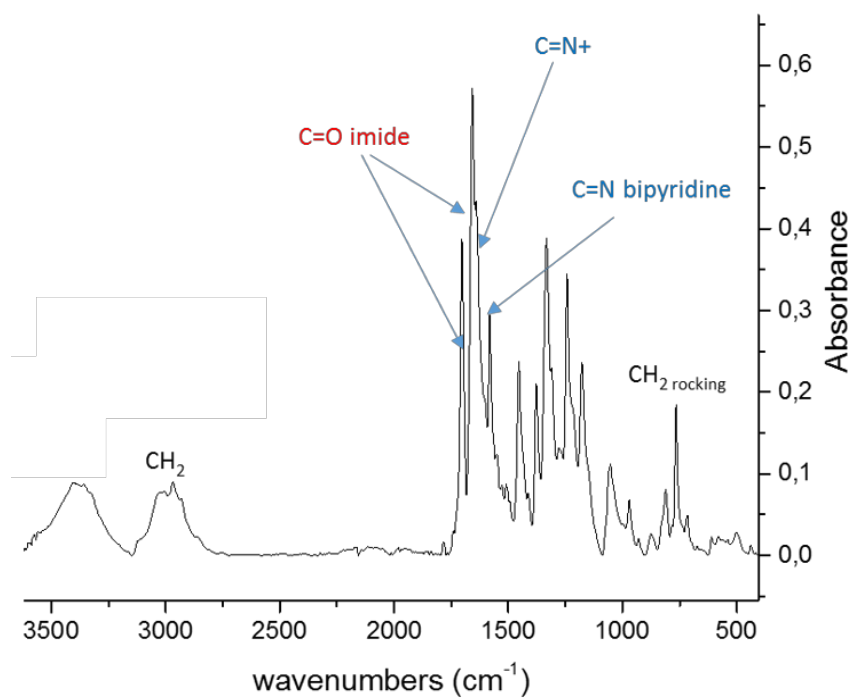

***Figure S3.*** *Typical FTIR-ATR spectrum of DNVBr.*

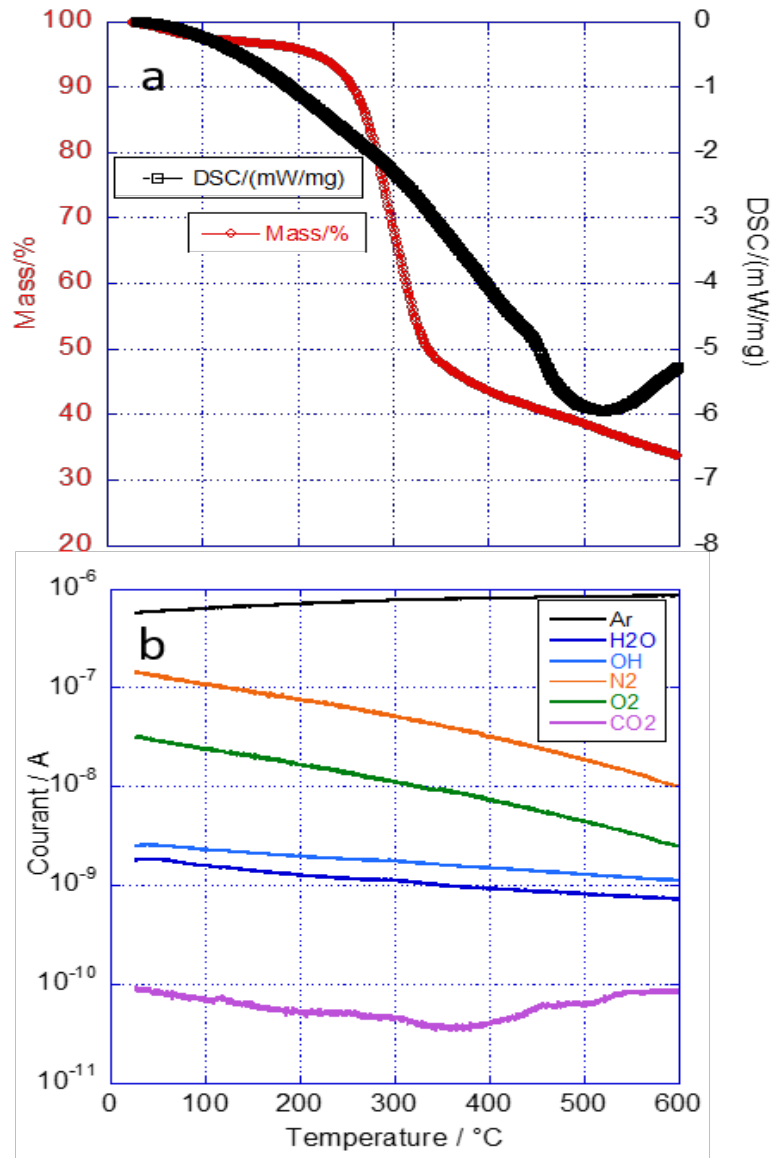

*Figure S4.* Thermal analysis data of DNVBr. (a) Typical TG (Mass%) and DSC (mW/mg) curves together with (b) corresponding gaz evolution as detected by means of mass spectrometry (MS) (m/z=17, is ascribed to OH, 18 to $H_2O$, 28 to $N_2$, 32 to $O_2$, 40 to Ar and 44 to $CO_2$) ; heating rate: 5°C/min under Ar.

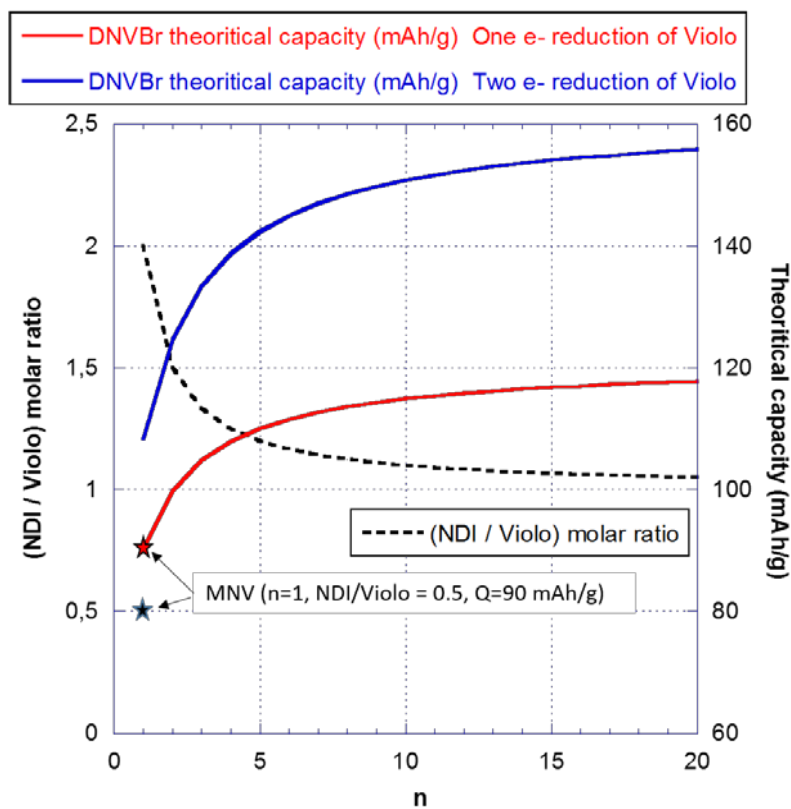

***Figure S5.*** *Evolution of the NDI/Violo molar ratio (black broken line) along with the corresponding theoretical specific capacity of DNVBr (red line, for a one electron redox process of the Violo unit, and, blue line, considering a two electron redox process) according to the number of repeating unit n (see **Scheme 1**). These values are compared to that of MNV (n=1, see **Scheme 1**) for which the NDI/Violo ratio is represented by a black star and the theoretical specific capacity by a red one.*

|  | $E_1$ (V vs. SCE) | $E_2$ (V vs. SCE) | $E_3$ (V vs. SCE) |
|---|---|---|---|
| **DNVBr** | -0.415 | -0.569 | -0.682 |
| **MNV** | -0.445 | -0.547 | -0.742 |
| $E_{DNV-MNV}$ **(mV)** | **+30** | **-22** | **+60** |

***Table S1.*** *Potential of redox processes defined as $E_i = 1/2(E_i^{peak,Ox} + E_i^{peak,Red})$. The key potential shift is highlighted in red.*

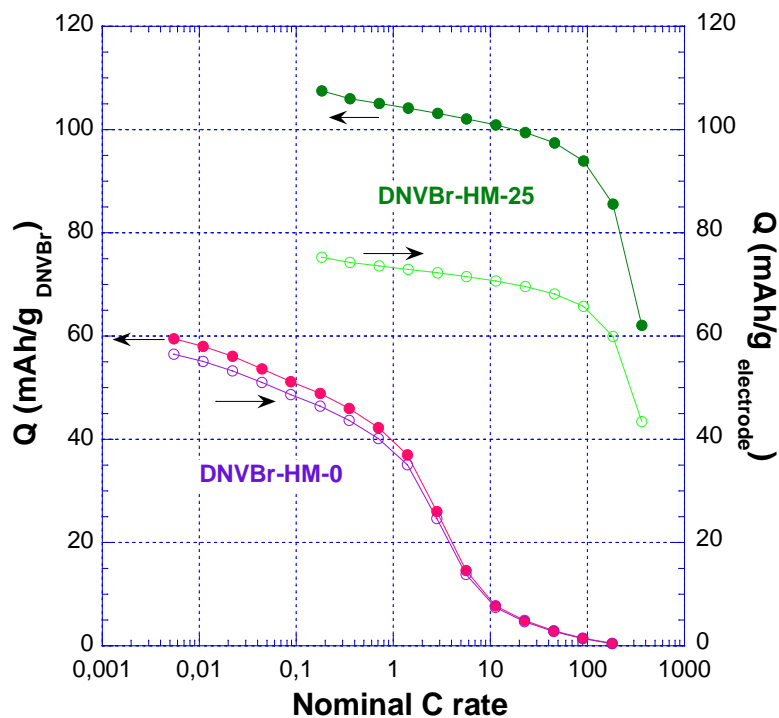

***Figure S6.*** *Ragone plot obtained in NaClO$_4$ 2.5 M for of DNVBr hand mixed with 25 wt% and 5 wt% of PTFE (referred to as DNVBr-HM-25) and with 5 wt% of PTFE but without carbon additive (referred to as DNVBr-HM-0). All capacity values are reported on oxidation (discharge of a full cell) by mass of DNVBr and by mass of the whole electrode. The extremely fast kinetics of DNVBr allows to obtain competitive capacity without carbon additive.*

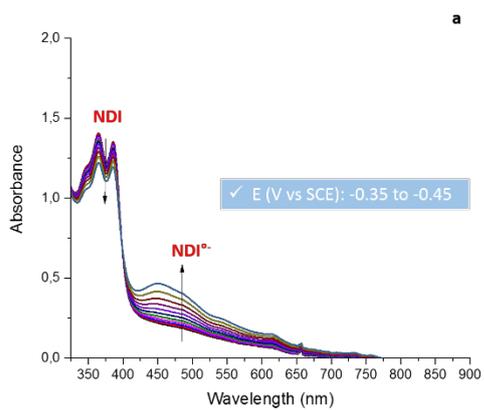
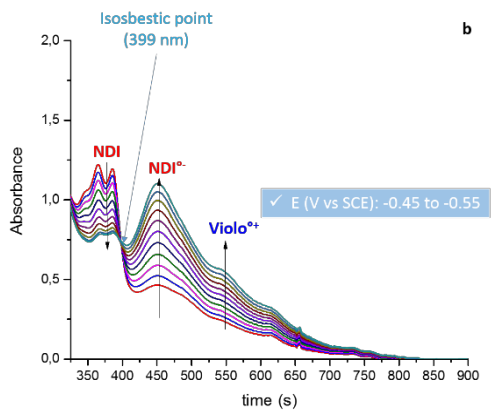
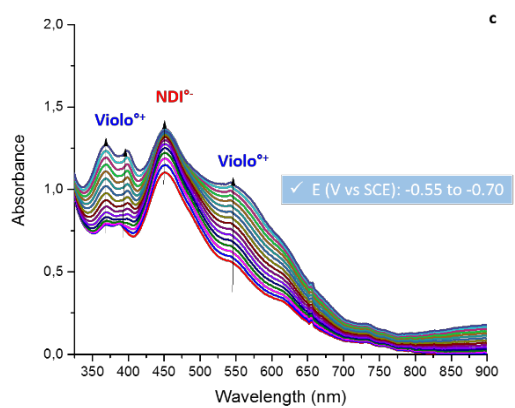
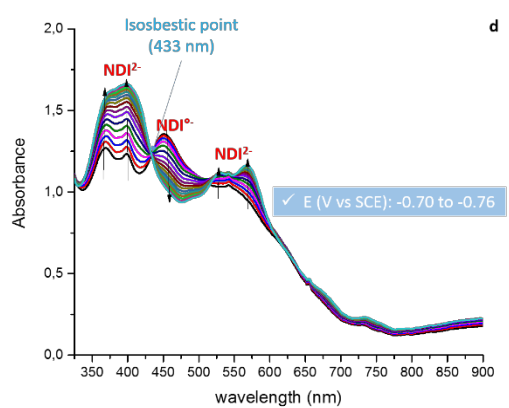
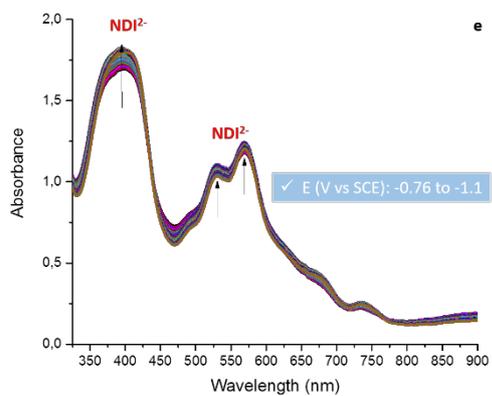
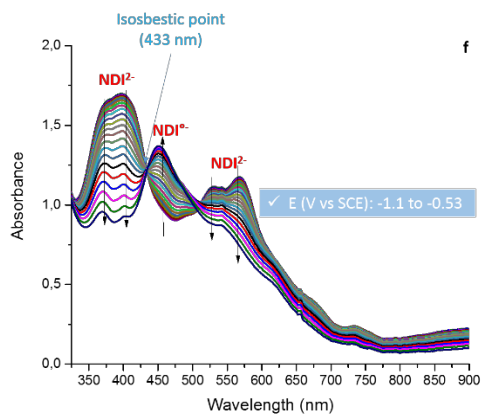

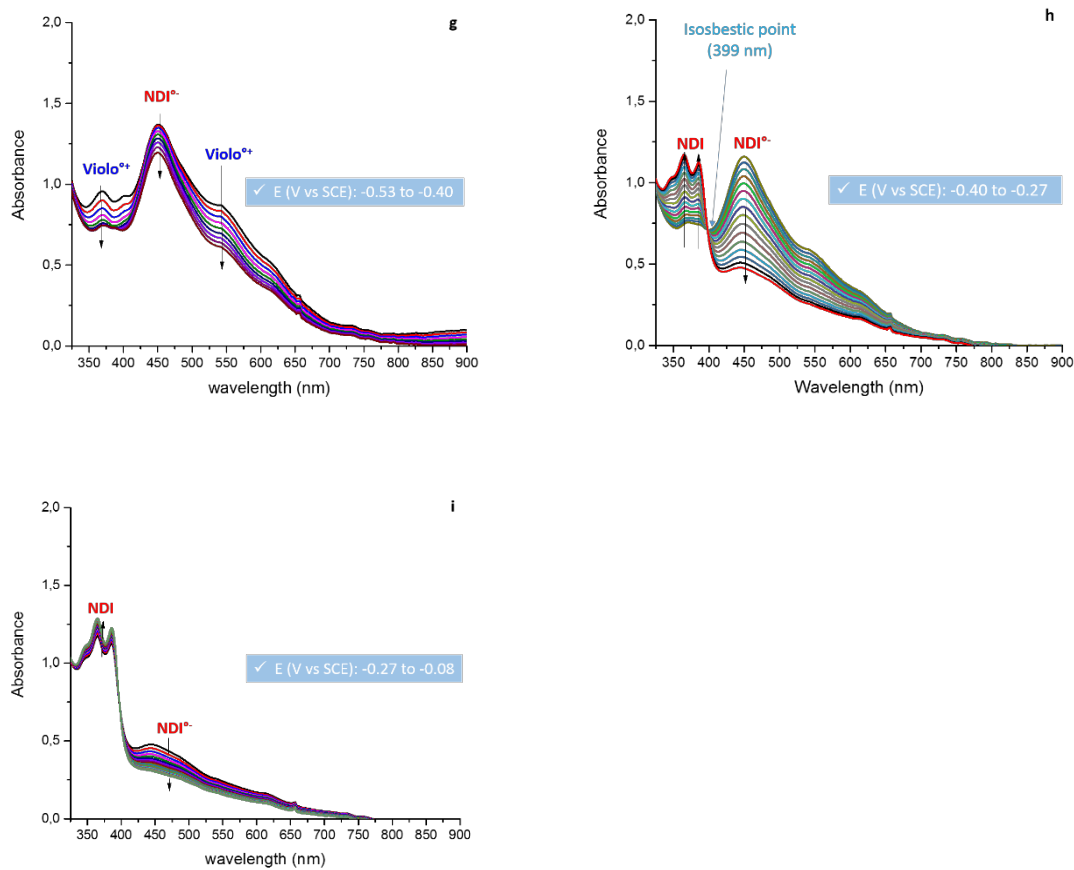

***Figure S7.*** *UV-vis spectroelectrochemical spectra showing the isosbestic points (b, d, f, h) characterizing the electrochemical behavior of DNVBr measured in NaClO$_4$ 2.5 M.*

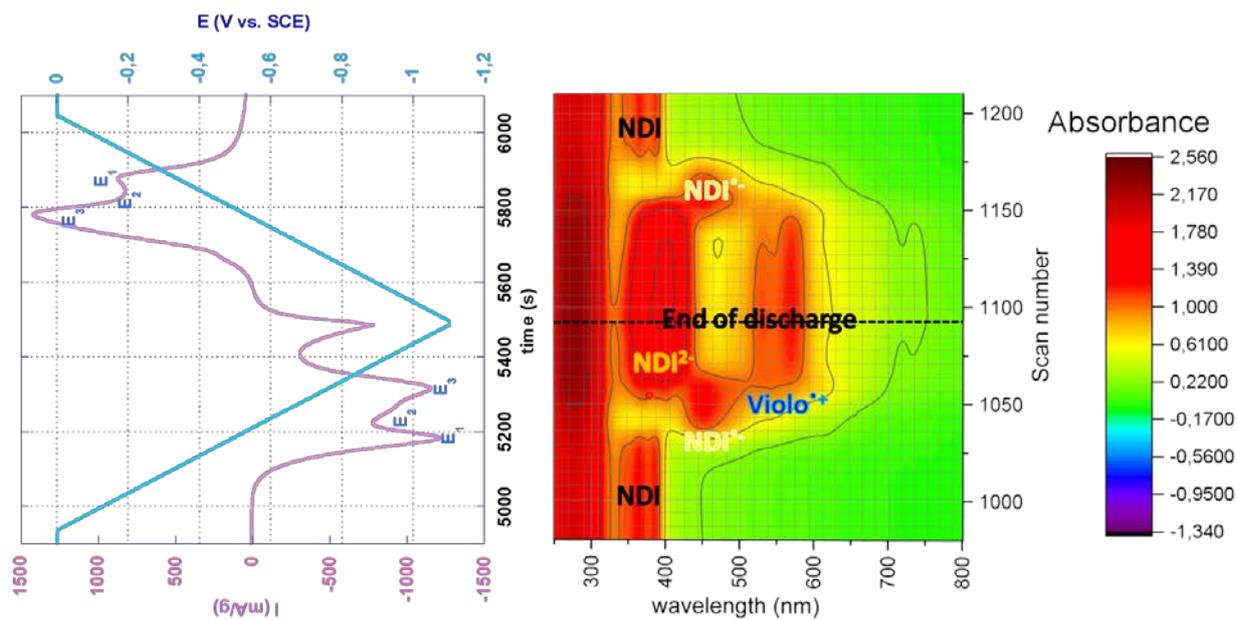

***Figure S8.*** *UV-vis spectroelectrochemical spectra and corresponding CV curves for DNVBr measured in Mg(ClO$_4$)$_2$ 1.25 M; scan rate: 2 mV/s.*

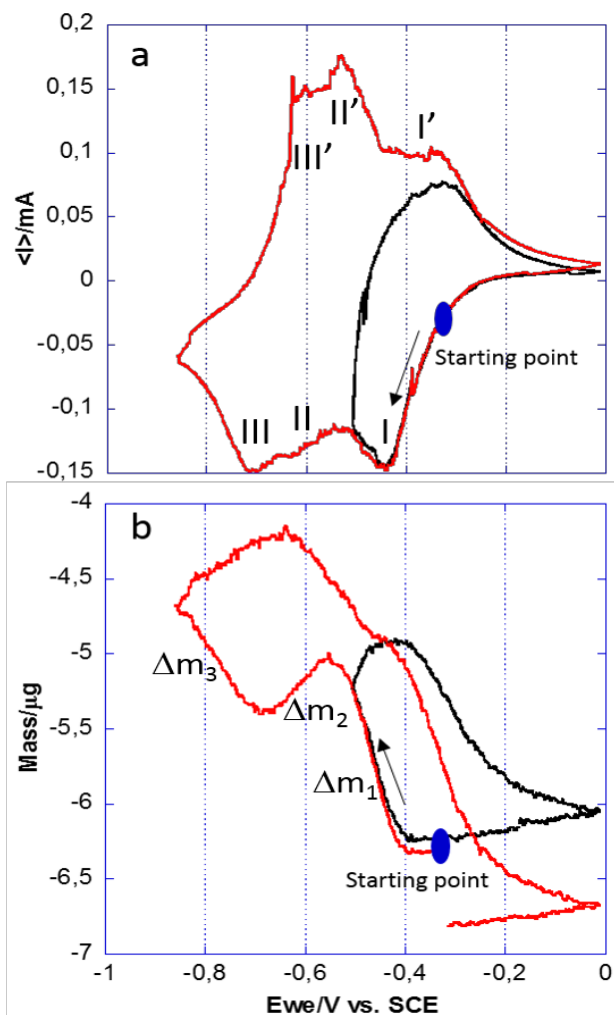

***Figure S9.*** *EQCM measurements obtained for DNVBr and measured in $Mg(ClO_4)_2$ 1.25 M with potential limitation to -0.5 V (black) and to -0.85 V (red), respectively; scan rate: 20 mV/s.*

In agreement with the UV-Vis spectroelectrochemistry, three successive steps are observed during the reduction: the reduction of DNVBr starts by a first gain of mass ($\Delta m_1$) that can be ascribed to the uptake of cationic species to compensate for formation of the radical anion ($NDI^{\bullet -}$) during (I). This process is followed by a mass loss ($\Delta m_2$) that is consistent with the reduction of the p-type Viologen to its cationic ($Violo^{\bullet +}$) form (II). Finally, the last step (III) consists in a second gain of mass ($\Delta m_3$) as expected considering the reduction of the radical anion ($NDI^{\bullet -}$) to its dianionic form $NDI^{2-}$. Therefore, the mass variations on reduction qualitatively matches the spectral changes observed by UV-Vis spectroelectrochemistry, supporting the dual cationic/anionic insertion processes. Quantification of these mass variations by determining the apparent molar mass $M_{app}$ defined as $M_{app} = (z \times F \times dm)/dQ$ where z is the ion charge, F the Faraday constant, using linear regression of the mass vs. charge plot[1]. The latter suggests an important co-insertion of water molecules during Mg cation ingress; 18 $H_2O$ molecules for (I) and 13 for (III). We note that a very similar result is obtained for (I) by limiting the potential to -0.5 V on reduction. During (II), results indicate departure of one $ClO_4^-$ ion is replaced by two water molecules. This value may not be accurate however, as (II) is intermixed with (III).

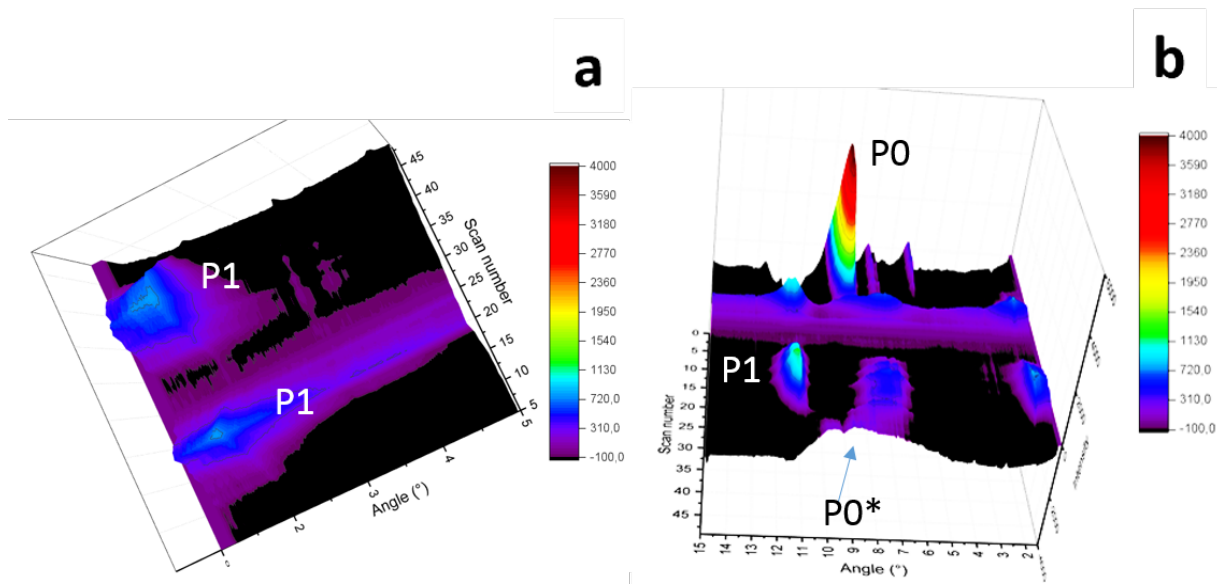

*Figure S10.* Operando XRD measurements of DNVBr during the first cycle.

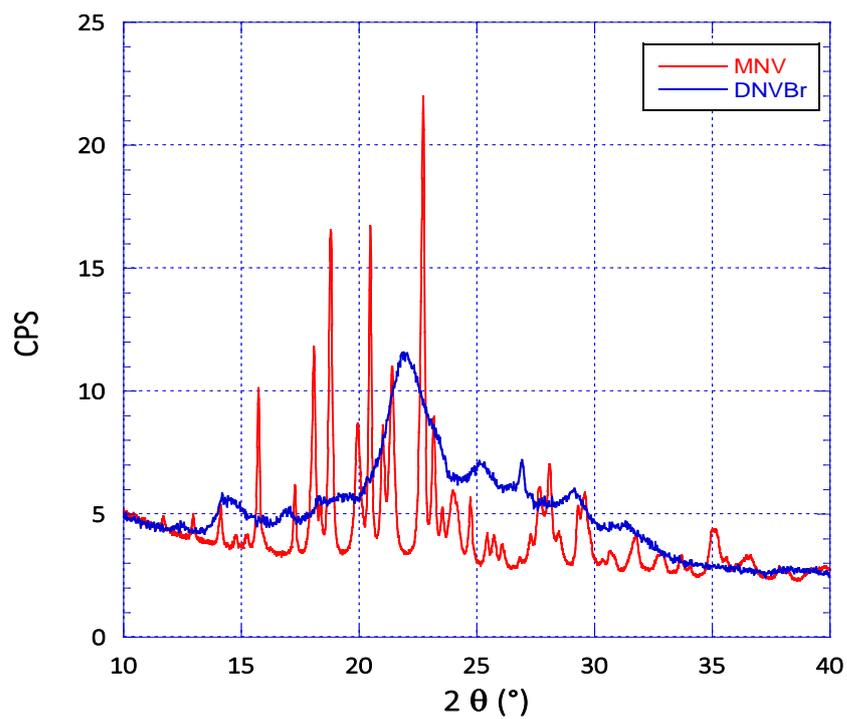

*Figure S11.* Powder XRD diagrams of MNV (red) and DNVBr (blue) ($\lambda$ = 1.5406 Å).

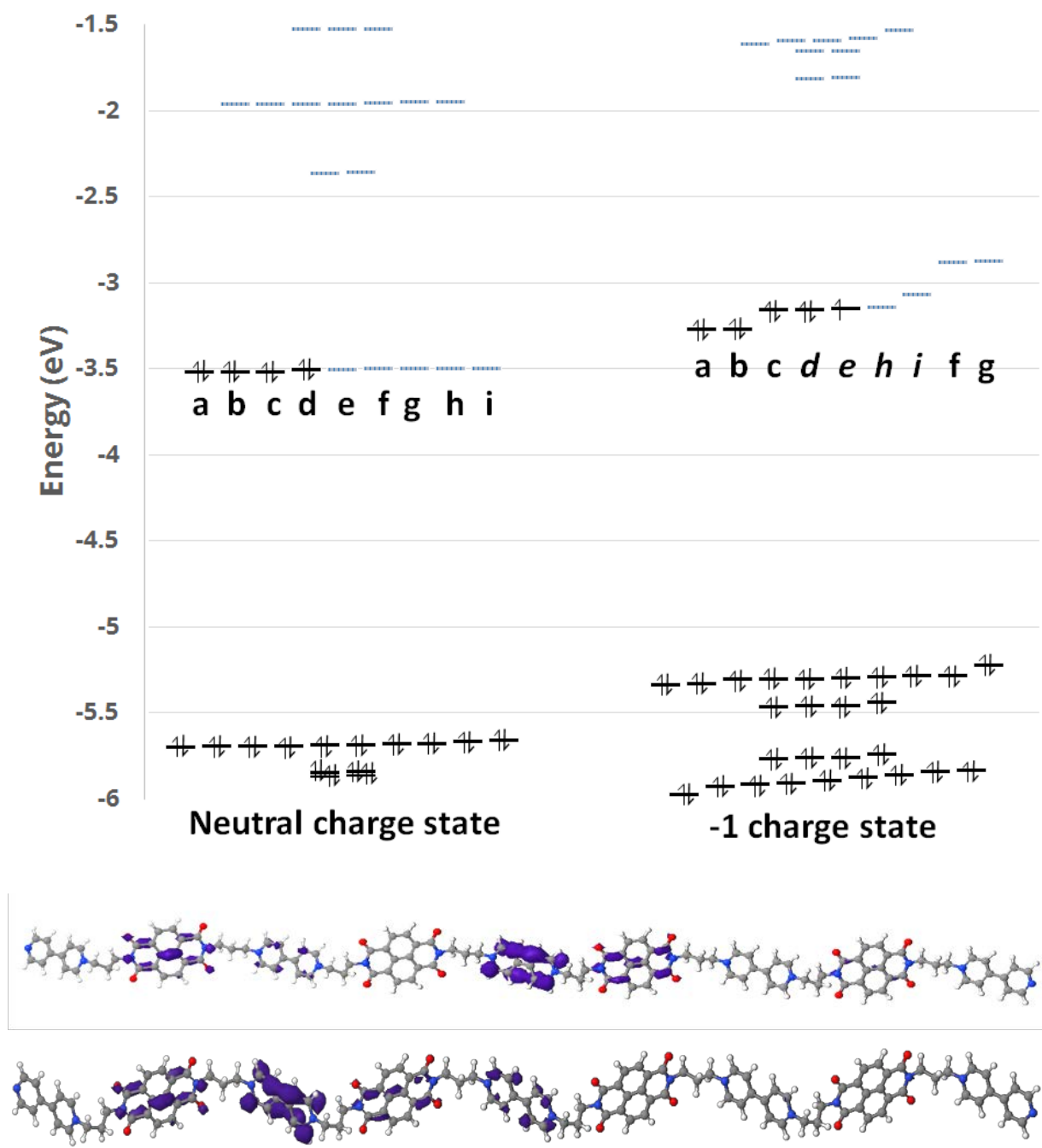

*Figure S12.* (up) Calculated LDA-DFT Kohn-Sham eigenvalues (eV) for an isolated DMV molecule in the (left) neutral and (right) -1 charge state. In -1 charge state a jellium background charge maintains unit cell neutrality, which contributes to the upward level shifts. Letters on left and right figures refer to electronic states as displayed in *Figure 4*, italics indicates the state is somewhat modified (for example sitting on only a single NDI group rather than a pair). (down) Examples of real-space wave function distribution on the molecule for two of the degenerated states at approx. -2 eV showing these empty states are more delocalized between Violo and NDI groups.

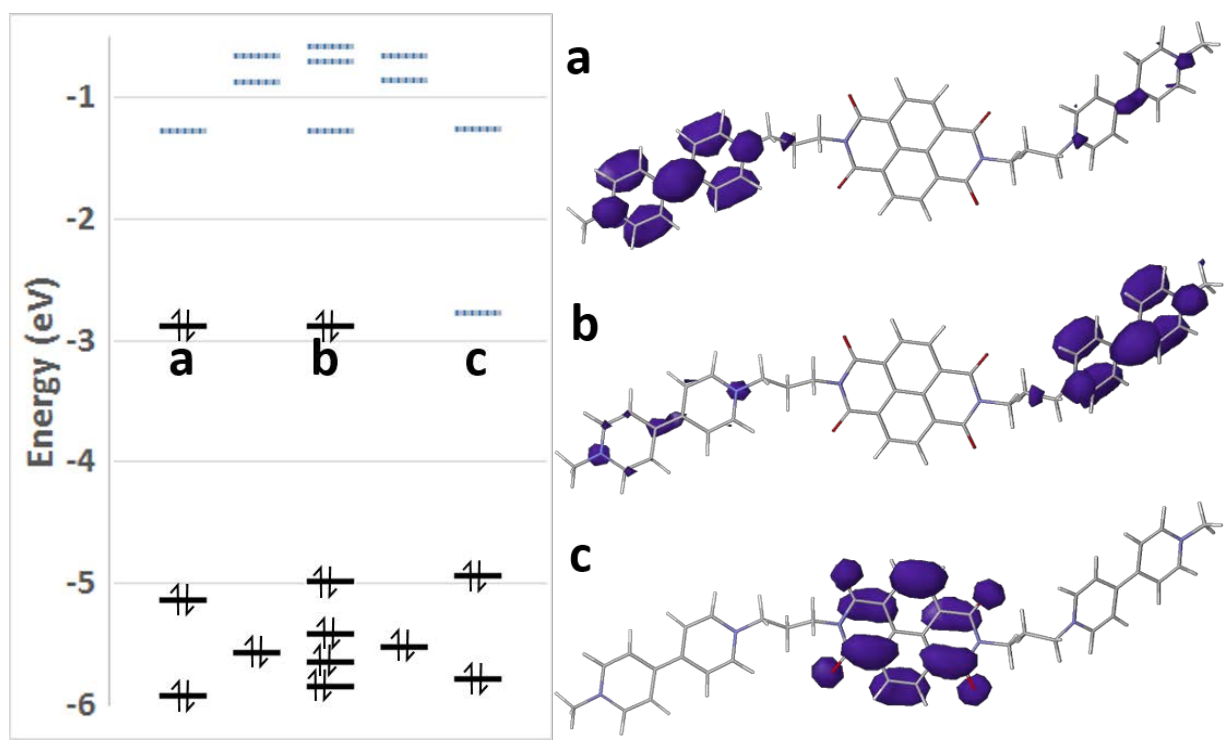

*Figure S13.* Calculated LDA-DFT Kohn-Sham eigenvalues (eV) for an isolated MNV molecule, showing the real-space wave function distribution on the molecule for states at the Fermi level. Level offset is for visual clarity and does not represent symmetry equivalence. The molecule is in the neutral charge state, i.e. after charge transfer from nominal counter-ions, which are not included explicitly in the calculation.

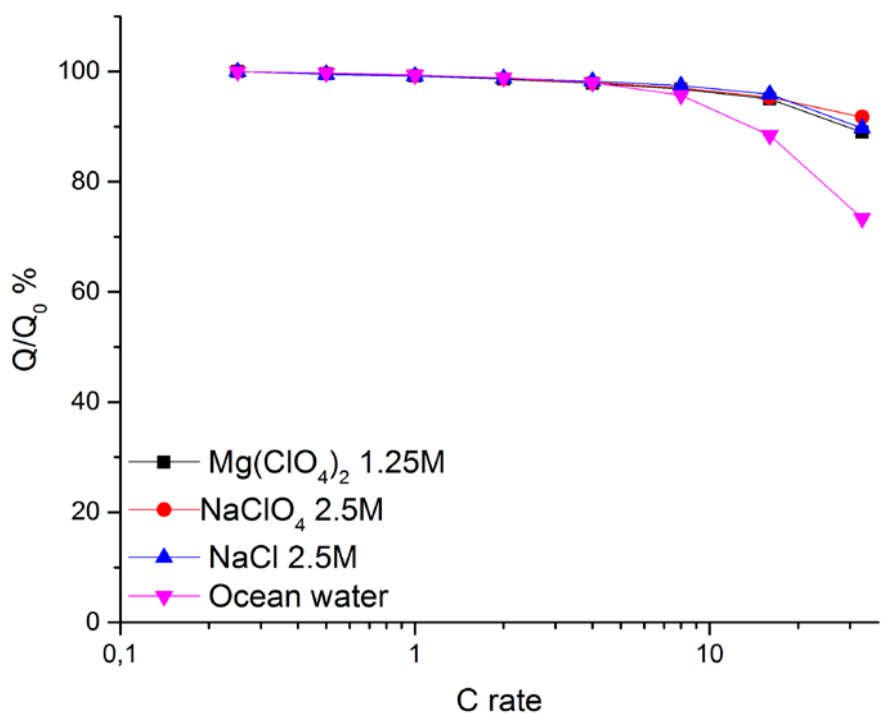

*Figure S14. Ragone plot of DNVBr measured in Mg(ClO$_4$)$_2$ 1.25 M (black squares), NaClO$_4$ 2.5 M (red dots), NaCl 2.5 M (blue triangles) and ocean water (pink triangles). Electrodes are in the 0.6-0.7 mAh/cm$^2$ range and contains 25 wt% Kb.*

| Electrolyte | Ionic conductivity at 25°C (S/m) |
|---|---|
| NaClO$_4$ 2.5 M | 12.8 |
| Mg(ClO$_4$)$_2$ 1.25 M | 12.3 |
| NaCl 2.5 M | 14.2 |
| Ocean water | 4.3 |

*Table S2. Conductivity of aqueous electrolytes at 25°C.*

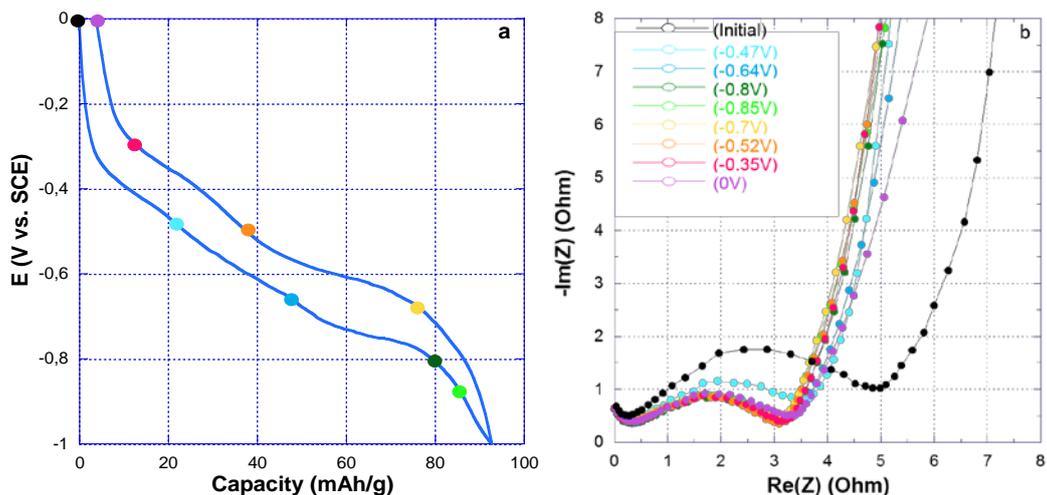

***Figure S15.*** *a) Galvanostatic cycling curve of DNVBr measured in NaClO$_4$ 2.5 M at 0.3 A/g (i.e., 4C rate) and, (b) Impedance (EIS) spectra of DNVBr measured at different potential values. For the sake of clarity, the potentials at which the EIS measurements have been performed are colored on the galvanostatic curve.*

*The relatively small decrease of resistance from the initial spectrum to the -0.47 V and -0.64 V ones, are tentatively ascribed to the improvement of contacts as a result of the impregnation and swelling of the electrode by the electrolyte.*

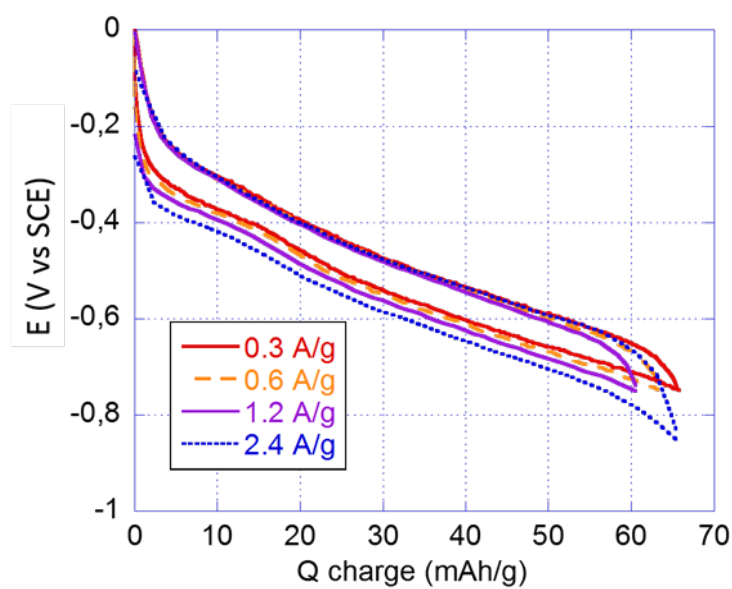

***Figure S16.*** *Galvanostatic cycling curve of DNVBr measured in NaClO$_4$ 2.5 M at different rates. A relatively small polarization is observed (especially on oxidation) upon increasing the current load by a factor 8, from 0.3 A/g (4C) to 2.4 A/g (32C).*

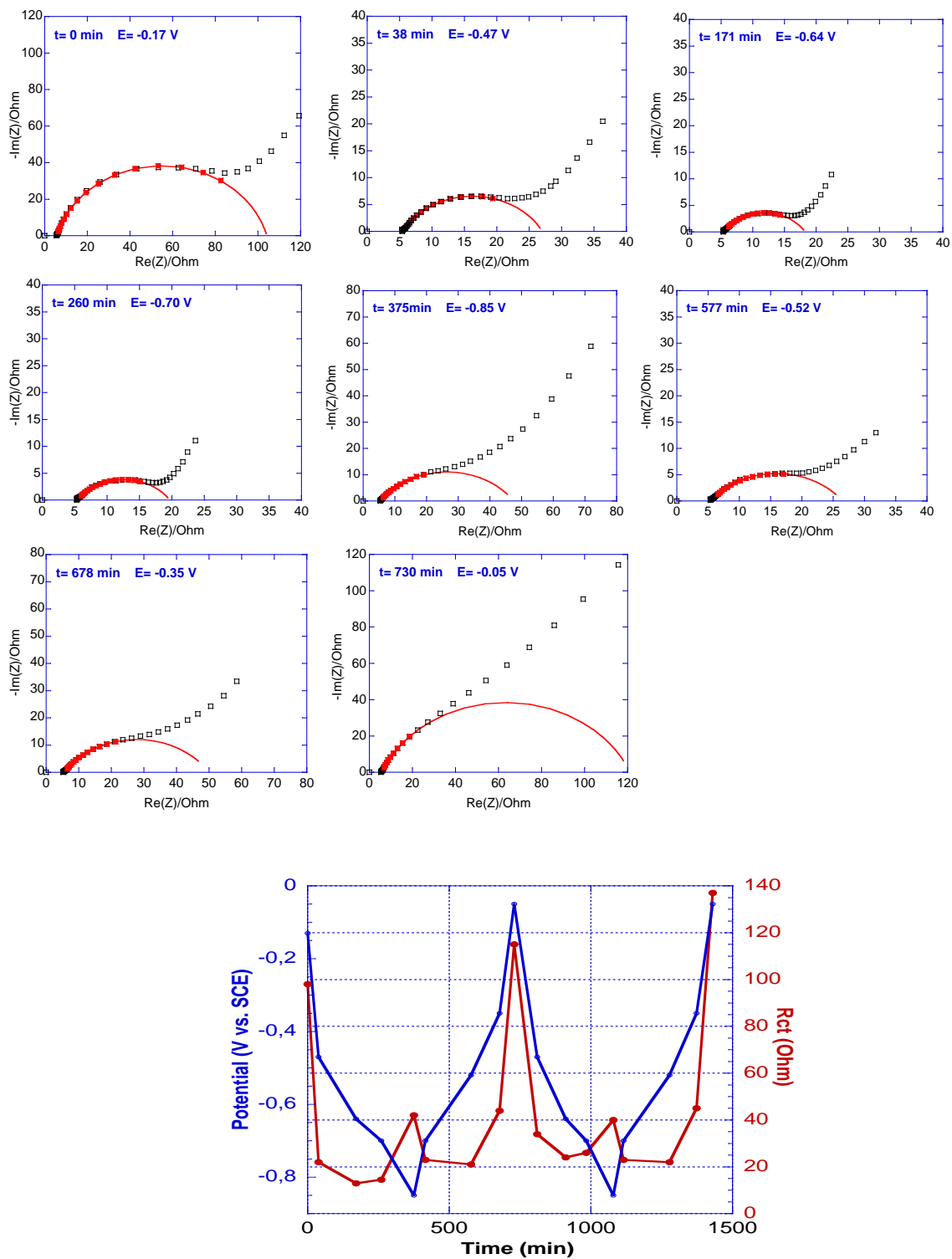

*Figure S17.* EIS spectra of DNVBr-HM-10 during the first cycle and fits of the semicircle using a constant phase element ($Q_2$) based equivalent circuit $R_1+Q_2/R_2$, $R_1$ being in the range of 5.5 ± 0.2 Ω. The characteristic frequency of each semicircle lies in between 10 and 21 Hz. The resulting evolution of the charge transfer resistance is shown during the first and second cycles along with potentials at which the EIS measurements have been performed.

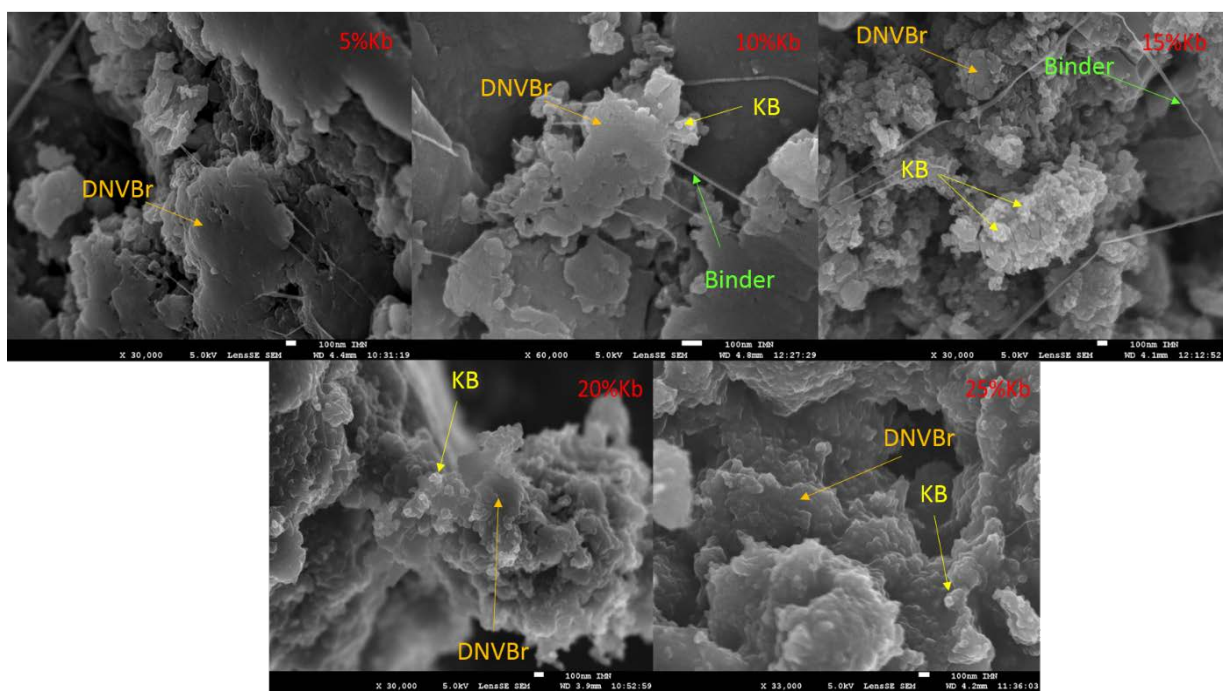

***Figure S18.*** *SEM images of DNVBr-HM-25% as well as DNVBr-BM-X% with X=5, 10, 15 and 20. Identification of each component was deduced by comparison using different values of X. The PTFE binder shows a fiber morphology as reported in the literature.[2]*

Note that SEM images of the samples provided in **Figure S18** distinctively show the three components of the composite electrodes. From 15 to 25 wt% of carbon additive, DNVBr particles are nicely wrapped by nano-grains of the carbon additive. Below 15wt%, however, carbon particles are sparsely distributed while the habitus of DNVBr grains shows flat faces, presumably as a result of the ball milling step. Complementary images taken at lower magnifications (**Figure S19**) reiterate these observations and highlight DNVBr grains are a few microns large.

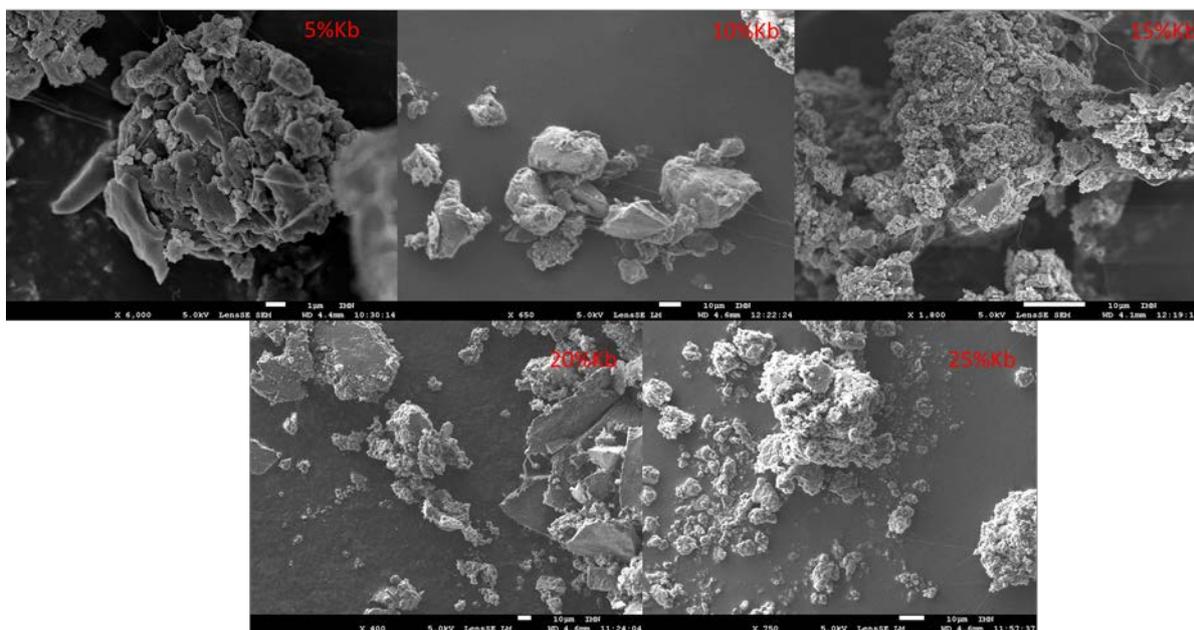

***Figure S19.*** *SEM images of DNVBr-HM-25% as well as DNVBr-BM-X% with X=5, 10, 15 and 20 at relatively low magnification.*

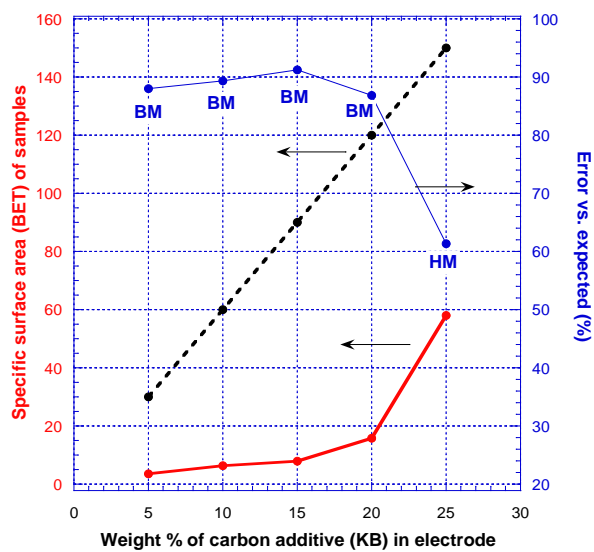

*Figure S20.* Estimated specific surface area (black) and experimental specific surface area normalized to the carbon content (red), for DNVBr-HM-25% and DNVBr-BM-X% with X=5, 10, 15 and 20. The error between experimental and estimated values is shown in blue. The theoretical specific surface area of each sample is dominated by that of the carbon additive (approx. 600 $m^2$/g). On this basis an estimation of the specific surface area for each sample is given by taking into account the respective carbon weight fractions (black). The experimental values derived from gas adsorption isotherms using the BET theory are however far below (red), this effect being even more pronounced for BM samples (larger errors, in blue).

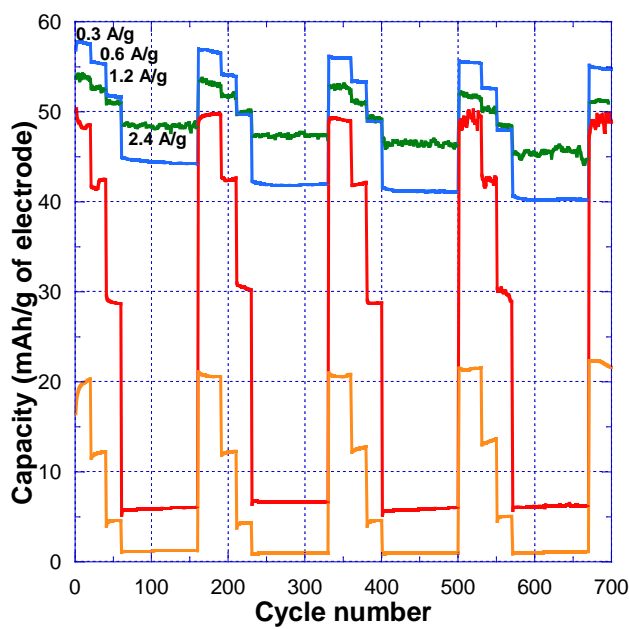

*Figure S21.* Cyclability curves (0; 0.85 V vs SCE as potential window) of DNVBr electrodes with different percentage of carbon measured in NaClO$_4$ 2.5 M: 20 wt% Kb BM (green), 15 wt% Kb BM (blue), 10 wt% Kb BM (red) and 5 wt% Kb BM (orange). Note that specific capacities refer to the mass of the whole electrode.

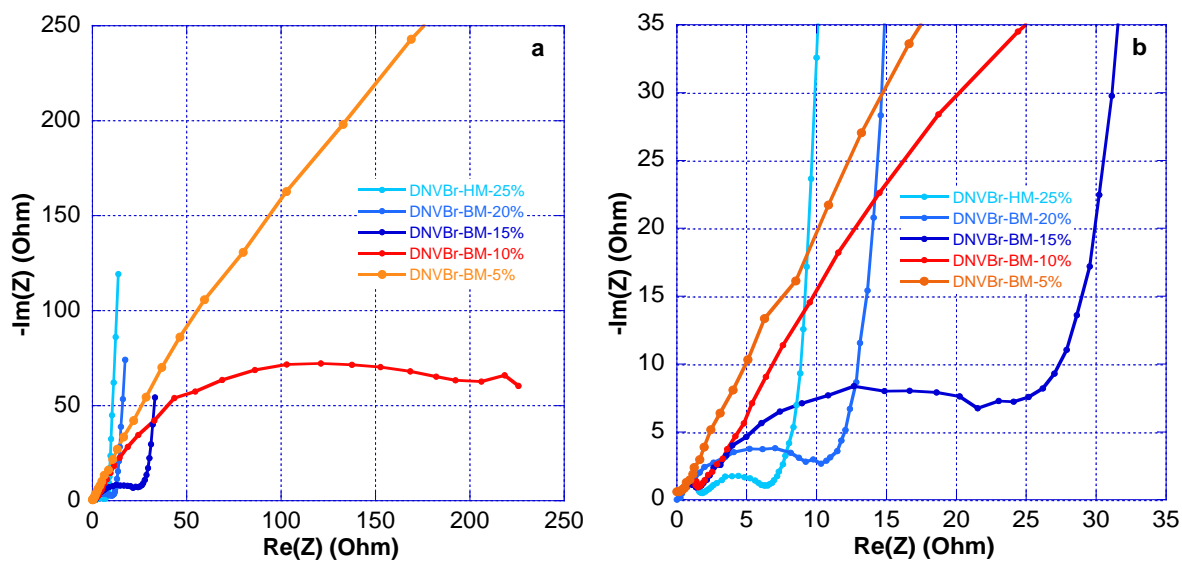

*Figure S22. a) EIS spectra of DNVBr electrodes prepared hand mixing with 25 wt% of carbon additive (light blue) and by ball milling with different mass fraction of carbon additive: 20 wt% Kb (blue), 15 wt% Kb (dark blue), 10 wt% Kb (red), 5 wt.% Kb (orange); b) Closest view of the spectra shown in a. All measurements have been performed in the initial state.*

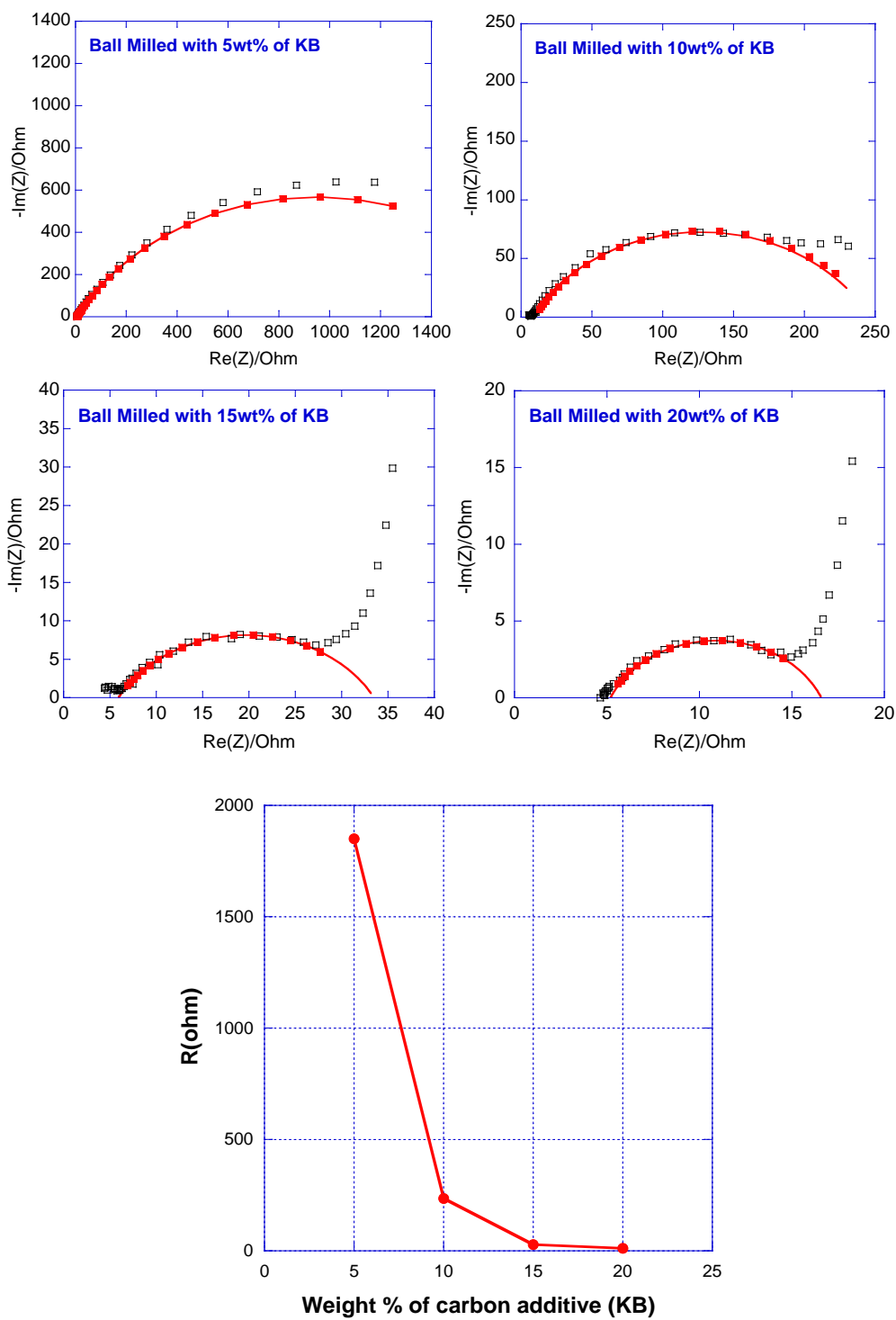

*Figure S23.* EIS spectra for DNVBr-BM-X% with X=5, 10, 15 and 20 in the initial state and fits of the semicircle using a constant phase element ($Q_2$) based equivalent circuit $R_1+Q_2/R_2$, $R_1$ being in the range of 5.4 ± 0.1 Ω. The characteristic frequency of each semicircle lies in between 18 and 23 Hz. The resulting evolution of the charge transfer resistance is shown according to the weight fraction of carbon additive (Kb).

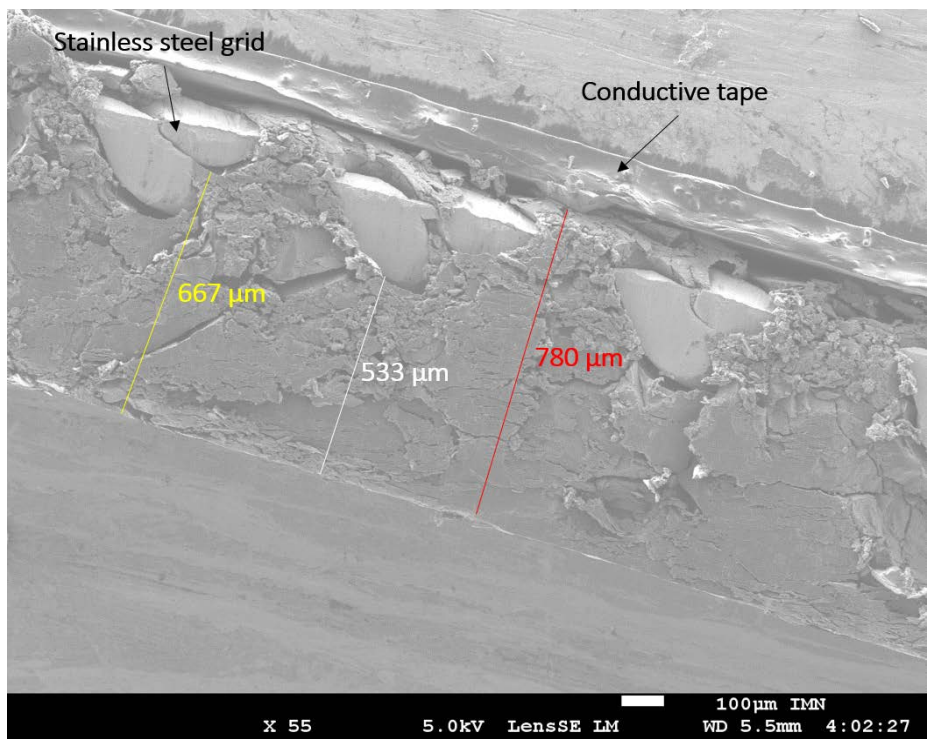

***Figure S24.*** *SEM cross section image of a DNVBr-HM-25 thick electrode (8 mAh/cm$^2$) pressed at 5 tons on a stainless steel grid (76.5 mg/cm$^2$ of DNVBr). The different thickness values are reported in the figure (0.53< thickness < 0.78 mm).*

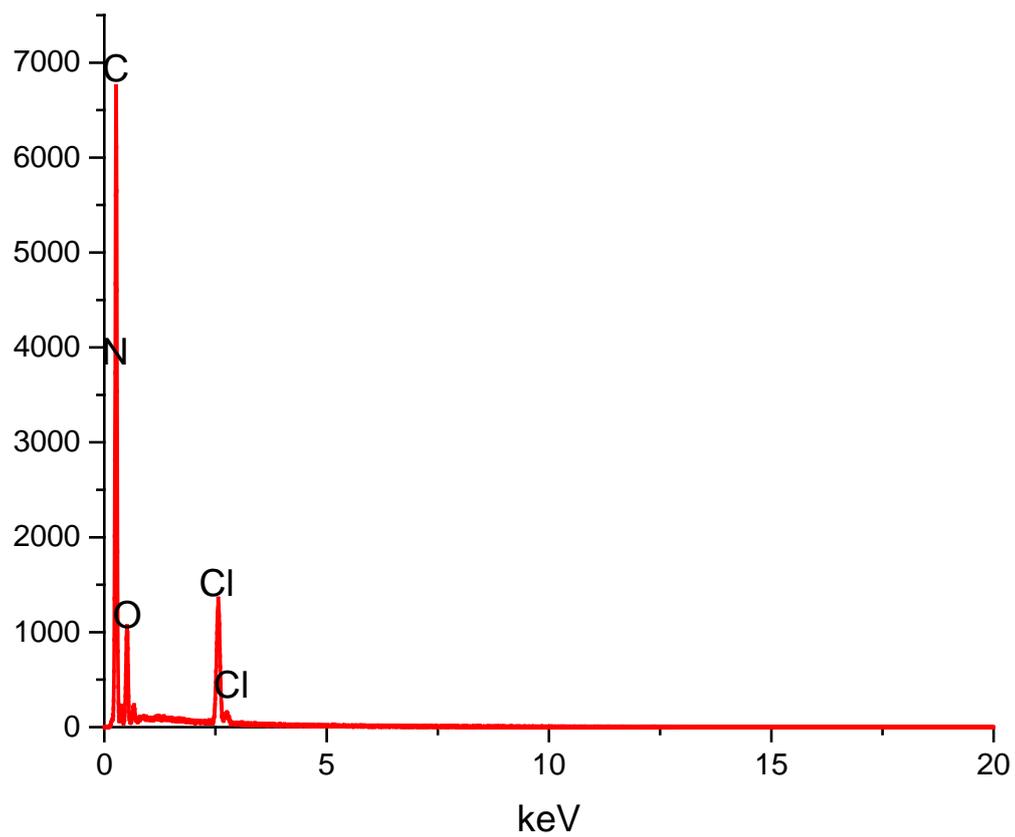

*Figure S25.* Typical EDS spectrum of DNVCl.

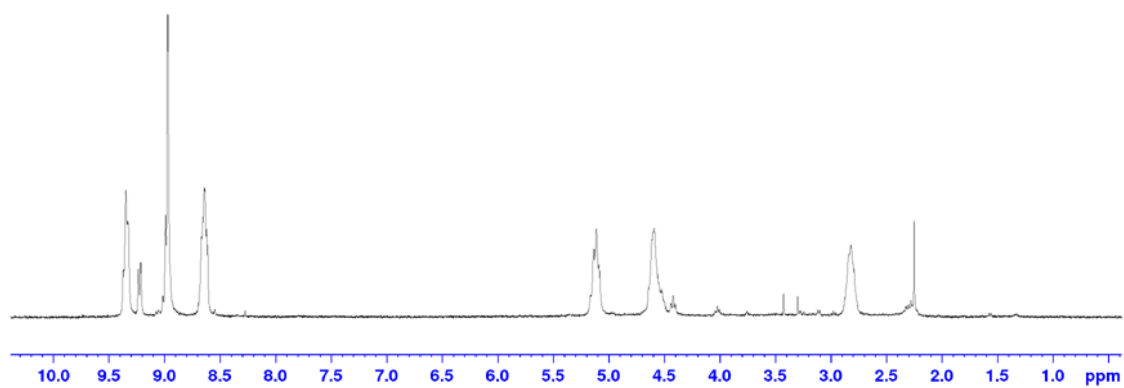

*Figure S26.* $^1$H NMR spectrum of DNVCl recorded in TFA-$d_1$.

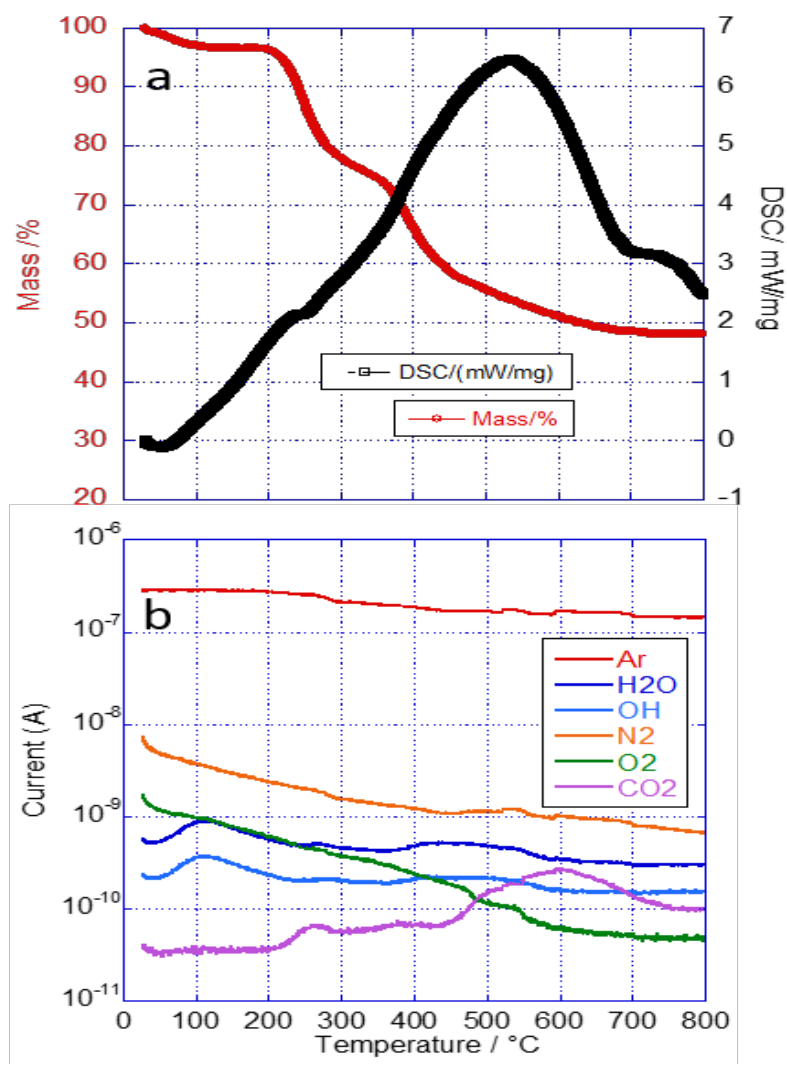

*Figure S27.* Thermal analysis data of DNVCl. (a) Typical TG (Mass%) and DSC (mW/mg) curves together with (b) corresponding gaz evolution as detected by means of mass spectrometry (MS) (m/z=17, is ascribed to OH, 18 to $H_2O$, 28 to $N_2$, 32 to $O_2$, 40 to Ar and 44 to $CO_2$) ; heating rate: 5°C/min under Ar.